# Statistical Inverse Problem

Yu. I. Bogdanov

*OAO "Angstrem", Moscow, Russia*
*E-mail: bogdanov@angstrem.ru*

November 27, 2002

> "Get at the root of it!"
> Koz'ma Prutkov

**Abstract**
A fundamental problem of statistical data analysis, density estimation by experimental data, is considered. A new method with optimal asymptotic behavior, the root density estimator, is proposed to solve the problem. The method is based on the representation of the probability density as a squared absolute value of a certain function, which is referred to as a psi function in analogy with quantum mechanics. The psi function is represented by an expansion in terms of an orthonormal set of functions. The expansion coefficients are estimated by the maximum likelihood method. An iteration algorithm for solving the likelihood equation is presented. The stability and rate of convergence of the solution are studied. A special iteration parameter is introduced: its optimal value is chosen on the basis of the maximin strategy. Numerical simulation is performed using the set of the Chebyshev—Hermite functions as a basis. It is shown that the introduction of the psi function allows one to represent the Fisher information matrix as well as statistical properties of the estimator of the state vector (state estimator) in simple analytical forms. A new statistical characteristic, a confidence cone, is introduced instead of a standard confidence interval. The chi-square test is considered to test the hypotheses that the estimated vector converges to the state vector of a general population and that both samples are homogeneous. The problem of choosing an optimal number of harmonics in the expansion is discussed. The method proposed may be applied to its full extent to solve the statistical inverse problem of quantum mechanics, namely, estimating the psi function on the basis of the results of mutually complementing experiments. The maximum likelihood technique and likelihood equation are generalized in order to analyze quantum mechanical experiments. The Fisher information matrix and covariance matrix are considered for a quantum statistical ensemble. The constraints on the energy are shown to result in high-frequency noise reduction in the reconstructed state vector.

**Introduction**

A key problem of statistical data analysis is the problem of estimating the probability distribution density. Almost all problems related to experimental data processing are reduced to either estimating the probability density (when experimental data have to be described in terms of statistical distributions) or determining the goodness of fit between data observed experimentally and theoretical density model (if exists).

Such a statement of the fundamental problem is recognized only de jure in the literature on mathematical statistics (and even not always). De facto, classical objects of mathematical statistics are smooth parametrized families of densities $p(x|\theta_1, \theta_2, ...)$ with either one or two unknown parameters $\theta_1, \theta_2, ...$ to be estimated using the observed data. The functional form of the density is assumed to be initially prescribed. Such a parametric analysis is well developed to estimate the parameters of a rather small number of distributions (including Gaussian, exponential, binomial, Poisson and several other specific distributions). The maximum likelihood method is regarded as the most perfect one to estimate the parameters. This method yields estimators that are close, in a certain sense, to the best possible estimators (see below).

The basic limitation of the traditional parametric approach is that it is impossible to describe distributions of an arbitrary form. This drawback has objective underlying causes. Indeed, the problem of statistical density estimator is an inverse problem of probability theory (while the direct problems are calculating various frequency quantities on the basis of a given model of a random event). The problem under consideration turns out to be ill-posed. This implies that in the absence of any a priori information about the distribution law $p(x)$ of a random variable, the problem of density estimation does not admit a solution.

The ill-posedness is a common feature of inverse problems. In the absence of additional information based on either objective knowledge or, at least, common sense (i.e., when there is no any a priori information), a researcher can try to seek a correct dependence in a wide class of functions. In this case, empirical data are sometimes insufficient to reliably estimate the statistical distribution, since there are a lot of functions that essentially differ from each other and, at the same time, correctly describe statistical data. Additional a priori considerations resulting in narrowing the class of functions are related to ranging solutions in their complexity. For example, one may consider lower harmonics as simpler compared to higher in standard sets of basis functions, or introduce so-called smoothing functionals etc.

A general approach to ill-posed problems was developed by Tikhonov [1]. An interpretation of an inverse problem of probability theory as ill-posed was given in the Prokhorov—Chentsov theory of probability measures (see Appendix 2 in [2]). Regularization of the problem of probability density estimation by smoothing an empirical distribution function is presented by Vapnik and Stefanyuk [3, 4]. Sometimes, it is convenient to smooth quantities found from an empirical distribution function by monotonous transformations rather than an empirical distribution itself. For instance, in the reliability assurance problems, it is convenient to perform smoothing in the so-called Weibull coordinates [5, 6].

In mathematical statistics, two basic kinds of estimators are usually considered: the kernel density estimator and orthogonal series estimator.

Kernel density estimators [7—10] (also called the Rosenblatt—Parzen estimators) are based on smoothing each point in a sample over its certain neighborhood. In this case, the density has the form:

$$\hat{p}(x) = \frac{1}{nh_n} \sum_{k=1}^{n} K\left(\frac{x - x_k}{h_n}\right), \quad (I.1)$$

where $x_1, \ldots, x_n$ is the sample of the size $n$; $K(x)$, the distribution density; and $h_n$, the sequence of the parameters describing the bandwidth.

If $n \to \infty$ and $h_n \to 0$, $\frac{1}{nh_n} \to 0$, under certain sufficiently general conditions, the kernel density estimator approaches the true density characterizing the general population. Corresponding density estimator is asymptotically unbiased, consistent, and asymptotically normal. A number of papers were devoted to the optimal selection of the bandwidth that is of a primary importance for kernel density estimation (see, e.g., [11—14]).

Now, the development of the theory of kernel density estimators is concerned with performing estimations in spaces of arbitrary nature allowing one to consider nonnumeric objects [15, 16].

The orthogonal series estimator [17—20] proposed by Chentsov (1962) is based on the expansion of an unknown distribution density into the Fourier series, and subsequent estimation of the expansion coefficients by a sample. The density estimator by first $m$ terms of the Fourier series is

$$\hat{p}(x) = \sum_{j=1}^{m} c_j \varphi_j(x), \quad (I.2)$$

where $c_j = \int \varphi_j^*(x)\hat{p}(x)dx \approx \frac{1}{n}\sum_{k=1}^{n}\varphi_j^*(x_k)$ (I.3)

and $\varphi_j(x), \quad j = 1, 2, \ldots$ is the orthonormal basis.

Both kernel and orthogonal series estimators can be regarded as delta sequence estimators [21].

The sequence of functions $\delta_m(x, y)$ of two arguments is referred to as a delta sequence if the integrals of these functions multiplied by arbitrary sufficiently good (for example, finite and infinitely differentiable) function $f(x)$ satisfy the condition

$$\lim_{m \to \infty} \int \delta_m(x, y) f(y) dy = f(x) \quad (I.4)$$

A delta-like character of kernel density estimators (at any finite $n$ and $h \to 0$) is evident directly.

For orthogonal series estimators, from (I.1)—(I.3) we find

$$\hat{p}(x) \approx \frac{1}{n}\sum_{k=1}^{n}\left(\sum_{j=1}^{m}\varphi_j^*(x_k)\varphi_j(x)\right) = \frac{1}{n}\sum_{k=1}^{n}\delta(x_k, x), \quad (I.5)$$

where the delta-like kernel is

$$\delta(x_k, x) = \sum_{j=1}^{m}\varphi_j^*(x_k)\varphi_j(x), \quad (I.6)$$

The delta-like character of the kernel follows from the completeness of the set of basis functions:

$$\sum_{j=1}^{\infty}\varphi_j^*(x_k)\varphi_j(x) = \delta(x - x_k) \quad (I.7)$$

The delta-like character of the estimators makes it possible to illustrate the ill-posedness of the inverse problem of probability theory. For arbitrary given set of experimental data, having passed through a certain optimum value, the density estimator begins gradually falling apart with increasing number of terms $m$ in the case of an orthogonal series estimator (or with decreasing bandwidth $h$ in the case of a kernel estimator) turning finally into a set of sharp peaks related to sample points.

Note that a delta sequence (I.6) is not a priori nonnegative. This may result in appearing meaningless negative values of probability density. Kernel density estimators have no this drawback, since the kernel is chosen by a researcher and, hence, can always be selected nonnegative. However, in the theory of kernel density estimators, kernels that are not positively defined are sometimes used in order to decrease the bias of density estimator [15].

A certain advantage of orthogonal density estimators is that they can yield analytical density approximation. Anyway, some data reduction takes place if the consideration is restricted to a few terms in the Fourier series. At the same time, kernel density estimators, generally speaking, do not provide data reduction, since they require permanently storing of all the initial data.

An orthogonal density estimator may be referred to as the Gram-Charlier estimator [22, 23]. In the Gram-Charlier estimators, it is assumed that the density is determined by a reference density $p_0(x)$ (as a rule, the Gaussian distribution) in zero-order approximation. If this density is given, it is better to use the functions that are orthonormal with respect to the weight function $p_0(x)$ instead of the ordinary orthonormal set:

$$\int \varphi_i^*(x)\varphi_j(x)p_0(x)dx = \delta_{ij} \quad (I.8)$$

The Chentsov estimator corresponds to the case $p_0(x)=1$.

Note that a set of functions that are orthonormal with respect to a weight function can be derived from any given system of linearly independent functions by a standard orthogonalization procedure.

An unknown density estimator in the Gram-Charlier method is sought in the form of the expansion

$$\hat{p}(x) = p_0(x)\sum_{j=1}^{m} c_j \varphi_j(x), \quad (I.9)$$

where the expansion coefficients are estimated by

$$c_j = \int \varphi_j^*(x)\hat{p}(x)dx \approx \frac{1}{n}\sum_{k=1}^{n}\varphi_j^*(x_k). \quad (I.10)$$

This paper is based on a symbiosis of mathematical tools of quantum mechanics and the Fisher maximum likelihood principle in order to find nonparametric (or, more precisely, multiparametric) effective density estimators with most simple (and fundamental) statistical properties.

The method proposed, the root density estimator, is based on the representation of the density in the form of a squared absolute value of a certain function, which is referred to as a psi function in analogy with quantum mechanics.

The introduction of the psi function results in substantial reduction in the structure of both the Fisher information matrix and covariance matrix of estimators, making them independent of a basis, allows one to provide positive definiteness of the density and represent results in a simple analytical form.

The root density estimator being based on the maximum likelihood method has optimal asymptotic behavior in contrast to kernel and orthogonal series estimators.

The likelihood equation in the method of the root density estimator has a simple quasilinear structure and admits developing effective rapidly converging iteration procedure even in the case of multiparametric problems (for example, when the number or parameters to be estimated runs up to many tens or even hundreds). That is why the problem under consideration favorably differs from the other well-known problems solved by the maximum likelihood method when the complexity of numerical simulations rapidly increases and the stability of algorithms decreases with increasing number of parameters to be estimated.

Basic objects of the theory (state vectors, information and covariance matrices etc.) become simple geometrical objects in the Hilbert space that are invariant with respect to unitary (orthogonal) transformations.

## 1. Maximum Likelihood Method and Fisher Information Matrix

Let $x=(x_1,...,x_n)$ be a sample under consideration represented by $n$ independent observations from the same distribution $p(x|\theta)$. Here, $\theta$ is the distribution parameter (in general, vector valued).

The likelihood function is determined by the following product:

$$L(x|\theta) = \prod_{i=1}^{n} p(x_i|\theta). \quad (1.1)$$

The formula under consideration is the $n$-dimensional joint density of random distributions of the components of the vector $x = (x_1,...,x_n)$ interpreted as a set of independent random variables with the same distribution. But if $x = (x_1,...,x_n)$ is a certain realization (fixed sample), the likelihood function as a function of $\theta$ characterizes the likeliness of various values of the distribution parameter.

According to the maximum likelihood principle put forward by Fisher in 1912 [24] and developed in the twenties of the last century [25], the value $\hat{\theta}$ from the region of acceptability, where the likelihood function reaches its maximum value, should be taken as an estimation for $\theta$.

As a rule, it is more convenient to deal with the log likelihood function:

$$\ln L = \sum_{i=1}^{n} \ln p(x_i|\theta). \quad (1.2)$$

Both the log likelihood and likelihood functions have extrema at the same points due to the monotonicity of the logarithm function.

The necessary condition for the extremum of the log likelihood function is determined by the likelihood equation of the form

$$\frac{\partial \ln L}{\partial \theta} = 0. \quad (1.3)$$

If $\theta = (\theta_1,...,\theta_s)$ is an $s$-dimensional parameter vector, we have the set of the likelihood equations

$$\frac{\partial \ln L}{\partial \theta_i} = 0 \qquad i = 1,...,s \quad (1.4)$$

The basic result of the theory of maximum likelihood estimation is that under certain sufficiently general conditions, the likelihood equations have a solution $\hat{\theta} = (\hat{\theta}_1,...,\hat{\theta}_s)$ that is a consistent, asymptotically normal, and asymptotically efficient estimator of the parameter $\theta = (\theta_1,...,\theta_s)$ [22, 26-28].

Formally, the aforesaid may be expressed as

$$\hat{\theta} \sim N(\theta, I^{-1}(\theta)). \quad (1.5)$$

The last formula means that the estimator $\hat{\theta}$ is asymptotically (at large $n$) a random variable with a multidimensional normal distribution with the mean equal to the true value of the parameter $\theta$ and the covariance matrix equal to the inverse of the Fisher information matrix.

The Fisher information matrix elements are

$$I_{ij}(\theta) = n \cdot \int \frac{\partial \ln p(x|\theta)}{\partial \theta_i} \frac{\partial \ln p(x|\theta)}{\partial \theta_j} p(x|\theta) dx \quad (1.6)$$

The factor $n$ indicates that the Fisher information is additive (the information of a sample consists of the information in its points). At $n \to \infty$, the covariance matrix asymptotically tends to zero matrix, and, in particular, the variances of all the components become zero (consistency).

The fundamental significance of the Fisher information consists in its property to set the constraint on achievable (in principle) accuracy of statistical estimators. According to the Cramer-Rao inequality [28], the matrix $\Sigma(\hat{\theta}) - I^{-1}(\theta)$ is nonnegative for any unbiased estimator $\hat{\theta}$ of an unknown vector valued parameter $\theta$. Here, $\Sigma(\hat{\theta})$ is the covariance matrix for the estimator $\hat{\theta}$. The corresponding difference asymptotically tends to a zero matrix for the maximum likelihood estimators (asymptotic efficiency).

## 2. Psi Function and Likelihood Equation

A psi function considered further is a mathematical object of statistical data analysis. This function is introduced in the same way as in quantum mechanics (see, e.g., [29-31]) to drastically simplify statistical density estimators obtained by the maximum likelihood method.

The introduction of the psi function implies that the "square root" of the distribution function

$$p(x) = |\psi(x)|^2 \quad (2.1)$$

is considered instead of the distribution function itself.

Let the psi function depend on $s$ unknown parameters $c_0, c_1, \ldots, c_{s-1}$ (according to quantum mechanics, the basis functions are traditionally numbered from zero corresponding to the ground state). The parameters introduced are the coefficients of an expansion in terms of a set of basis functions:

$$\psi(x) = \sum_{i=0}^{s-1} c_i \varphi_i(x). \quad (2.2)$$

Assume that the set of the functions is orthonormal. Then, the normalization condition (the total probability is equal to unity) is given by

$$\sum_{i=0}^{s-1} c_i c_i^* = 1. \quad (2.3)$$

Hereafter, the asterisk denotes the complex conjugation.

We will consider sets of basis functions that are complete for $s \to \infty$. At a finite $s$, the estimation of the function by (2.2) involves certain error. The necessity to restrict the consideration to a finite number of terms is related to the ill-posedness of the problem. Because of the finite set of experimental data, a class of functions, where the psi function is sought, should not be too wide; otherwise a stable description of a statistical distribution would be impossible. Limiting the number of terms in the expansion by a finite number $s$ results in narrowing the class of functions where a solution is sought. On the other hand, if the class of functions turns out to be too narrow (at too small $s$), the dependence to be found would be estimated within too much error. The problem of an optimal choice of the number of terms in the expansion is discussed in greater detail in Sec. 7.

The maximum likelihood method implies that the values maximizing the likelihood function and its logarithm

$$\ln L = \sum_{k=1}^{n} \ln p(x_k|c) \to \max \quad (2.4)$$

are used as most likely estimators for unknown parameters $c_0, c_1, \ldots, c_{s-1}$.

The probability density is

$$p(x) = \psi^* \psi = c_i c_j^* \varphi_i(x) \varphi_j^*(x). \quad (2.5)$$

Hereafter, we imply the summation over recurring indices numbering the terms of the expansion in terms of basis functions (unless otherwise stated). On the contrary, statistical sums denoting the summation over the sample points will be written in an explicit form.

At sample points, the distribution density is

$$p(x_k) = \psi^* \psi = c_i c_j^* \varphi_i(x_k) \varphi_j^*(x_k). \quad (2.6)$$

In our case, the likelihood function has the form

$$\ln L = \sum_{k=1}^{n} \ln\left[ c_i c_j^* \varphi_i(x_k) \varphi_j^*(x_k) \right]. \quad (2.7)$$

In view of the normalization condition, seeking an extremum of the log likelihood function is reduced to that for the following function:

$$S = \ln L - \lambda\left( c_i c_i^* - 1 \right), \quad (2.8)$$

where $\lambda$ is the Lagrange multiplier.

The necessary condition for an extremum yields the likelihood equation

$$\frac{\partial S}{\partial c_i^*} = \sum_{k=1}^{n} \frac{\varphi_i^*(x_k) \varphi_j(x_k)}{p(x_k)} c_j - \lambda c_i = 0. \quad (2.9)$$

Thus, the problem of looking for the extremum is reduced to the eigenvalue problem

$$R_{ij} c_j = \lambda c_i \qquad i, j = 0, 1, \ldots, s-1, \quad (2.10)$$

where

$$R_{ij} = \sum_{k=1}^{n} \frac{\varphi_i^*(x_k) \varphi_j(x_k)}{p(x_k)}. \quad (2.11)$$

The problem (2.10) is formally linear. However, the matrix $R_{ij}$ depends on an unknown density $p(x)$. Therefore, the problem under consideration is actually nonlinear, and should be solved by the iteration method (see below).

An exception is the histogram density estimator presented below when the problem can be solved straightforwardly.

Multiplying both parts of Eq. (2.10) by $c_i^*$ and summing with respect to $i$, in view of (2.3) and (2.5), we find that the most likely state vector $c$ always corresponds to its eigenvalue

$$\lambda = n.$$

Let us verify whether the substitution of the true state vector into the likelihood equation turns it into an identical relation (in the asymptotic limit). Indeed, at a large sample size ($n \to \infty$), according to the law of large numbers (the sample mean tends to the population mean) and the orthonormality of basis functions, we have

$$\frac{1}{n}R_{ij} = \frac{1}{n}\sum_{k=1}^{n}\frac{\varphi_i^*(x_k)\varphi_j(x_k)}{p(x_k)} \rightarrow$$

$$\rightarrow \int \frac{\varphi_i^*(x)\varphi_j(x)}{p(x)}p(x)dx = \delta_{ij} \qquad (2.12)$$

Thus, the matrix $\frac{1}{n}R$ asymptotically tends to unit matrix. In other words, Eq. (2.10) shows that the true state vector is its solution for $n \rightarrow \infty$ (consistency). The matrix $\frac{1}{n}R$ may be referred to as a quasi-unit.

Let us assume that the basis functions $\varphi_i(x)$ and the state vector $c$ are real valued. Then, the basic equation for the state vector (2.10) can be expressed in the form

$$\frac{1}{n}\sum_{k=1}^{n}\left(\frac{\varphi_i(x_k)}{\sum_{j=0}^{s-1}c_j\varphi_j(x_k)}\right) = c_i \qquad i = 0,1,\ldots,s-1 \qquad (2.13)$$

Here, we have written the summation signs for clearness. As is easily seen, the solution of this equation satisfies the normalization condition (2.3).

3. **Histogram Density Estimator**

In order to study the histogram density estimator, one has to assume that a distribution is given in a finite region (in the case of variables distributed along an infinite interval, it is necessary to cut off the distribution tails, e.g., by using maximum and minimum values in the sample as bounds).

Let us divide the full range of variation for a random variable into a finite number of intervals. Points $x_0, x_1, \ldots, x_s$ divide the full range of variation for a random variable into $s$ intervals (bins).

Assume that

$$\varphi_i(x) = \begin{cases} \frac{1}{(x_{i+1}-x_i)^{1/2}} & \text{at} \quad x_i \leq x \leq x_{i+1} \\ 0 & \text{otherwise} \end{cases} \qquad (3.1)$$

The functions $\varphi_i(x) \quad i = 0,1,\ldots,s-1$ form an orthonormal but, of course, incomplete set.

Equation (2.3) yields the following most likely estimator for the psi function:

$$\psi(x) = \sum_{i=0}^{s-1}c_i\varphi_i(x), \qquad c_i = \sqrt{\frac{n_i}{n}}, \qquad (3.2)$$

where $n_i$ is the number of points in $i$-th interval.

In order to avoid appearing indeterminate forms (zero divided by zero) while calculating the expansion coefficients $c_i$, one has to assume that $n_i > 0$ in each interval.

As is easily seen, the square of the psi function constructed in this way is a histogram density estimator.

Applying a unitary transformation to the found state vector shows a natural way to smooth a histogram density estimator that is as follows.

Let us transform the column vector $c_i$ and basis functions $\varphi_i(x)$ by a unitary matrix $U$:

$$c'_i = U_{ik} c_k, \quad (3.3)$$

$$\varphi'_i(x) = U^*_{il} \varphi_l(x). \quad (3.4)$$

The psi function and, hence, the density turn out to be invariant with respect to this transformation. Indeed,

$$\psi'(x) = c'_i \varphi'_i(x) = U_{ik} c_k U^*_{il} \varphi_l(x) = c_l \varphi_l(x). \quad (3.5)$$

Here, we have taken into account that due to the unitarity of the matrix $U$,

$$U^*_{il} U_{ik} = U^+_{li} U_{ik} = \delta_{lk}. \quad (3.6)$$

The plus superscript denotes the Hermitian conjugation.

The aforementioned transformation will be useful if the basis functions $\varphi'_i(x)$ in a new representation may be ranged in increasing complexity in such a way that the amplitudes $c'_i$ corresponding to first (most simple) basis functions turn out to be large, whereas those corresponding to more complex basis functions, relatively small. Then, a histogram density can be smoothed by truncating higher harmonics.

A classical example of such a unitary transformation is the discrete Fourier transform given by the matrix

$$U_{kl} = \frac{1}{\sqrt{s}} \exp\left(i \frac{2\pi}{s} kl\right). \quad (3.7)$$

A state vector resulting from the unitary transformation with the matrix (3.7) may be interpreted as a frequency spectrum of a signal that is the histogram estimator of a psi function.

In general case, choosing a unitary transformation and a way in which to filter noise, and ranging basis functions in ascending order of complexity should be performed on the basis of the analysis of a particular problem. A statistical fluctuation level for an empirical psi function is discussed in Sec. 5.

## 4. Computational Approach to Solving Likelihood Equation

In order to develop an iteration procedure for Eq. (2.13), let us rewrite it in the form

$$\alpha c_i + (1-\alpha) \frac{1}{n} \sum_{k=1}^{n} \left( \frac{\varphi_i(x_k)}{\sum_{j=0}^{s-1} c_j \varphi_j(x_k)} \right) = c_i. \quad (4.1)$$

Here, we have introduced an additional parameter $0 < \alpha < 1$ that does not change the equation itself but substantially influences the solution stability and the rate of convergence of an iteration procedure.

Let us represent an iteration procedure (transition from $r$-th to $r+1$-th approximation) in the form

$$c_i^{r+1} = \alpha \cdot c_i^r + (1-\alpha) \cdot \frac{1}{n} \sum_{k=1}^{n} \left( \frac{\varphi_i(x_k)}{\sum_{j=0}^{s-1} c_j^r \varphi_j(x_k)} \right). \quad (4.2)$$

Let us study the conditions of stable convergence of the iteration procedure to the solution. We restrict our consideration to the case of small deviations. Let $\delta c$ be any small deviation of an approximate state vector from the exact solution of Eq. (4.1) at arbitrary step; and $\delta c'$, that at the next iteration step. The squared distance between the exact and approximate solutions is $(\delta c)^T (\delta c)$.

A fundamental condition for an iteration procedure to converge is that the corresponding mapping has to be contracting (see the principle of contracting mappings and fixed point theorem [32, 33]). A mapping is contracting if $(\delta c')^T (\delta c') < (\delta c)^T (\delta c)$. It can be proved that

$$\delta c' = A \delta c, \quad (4.3)$$

where $A$ is the perturbation matrix:

$$A = \alpha E - \frac{(1-\alpha)}{n} R. \quad (4.4)$$

Here, $E$ is an $(s \times s)$ unit matrix.
After an iteration, the squared distance is decreased by

$$(\delta c)^T (\delta c) - (\delta c')^T (\delta c') = (\delta c)^T B (\delta c), \quad (4.5)$$

where 
$$B = E - A^T A \quad (4.6)$$

is the contracting matrix.

The mapping is contracting if $B$ is a positive matrix.

The minimum eigenvalue $\lambda_{\min}$ of the $B$ matrix is expedient to consider as a measure of contractility. An eigenfunction related to $\lambda_{\min}$ corresponds to perturbation that is worst from the convergence standpoint. Thus, the parameter $\lambda_{\min}$ characterizes the guaranteed convergence, since the squared distance decreases at least by $\lambda_{\min} \cdot 100\%$ at each step.

Let $R_0$ be the vector of eigenvalues for the $R$ matrix.

The $B$-matrix eigenvalues are expressed in terms of the $R$-matrix eigenvalues by

$$\lambda_i = 1 - \alpha^2 + \frac{2\alpha(1-\alpha)}{n} R_{0i} - \frac{(1-\alpha)^2}{n^2} R_{0i}^2 \quad (4.7)$$

The minimum of this expression at any given $\alpha$ is determined by either maximum (at small $\alpha$) or minimum (at large $\alpha$) $R_{0i}$.

As an optimal value of $\alpha$, we will use the value at which $\lambda_{\min}$ reaches its maximum (maximin rule). An optimal value of $\alpha$ is determined by the sum of maximum and minimum values of $R_{0i}$:

$$\alpha_{opt} = \frac{D_0}{2n + D_0}, \quad D_0 = \max(R_{0i}) + \min(R_{0i}). \quad (4.8)$$

For the difference of distances between the approximate and exact solutions before and after iteration, we have

$$\rho' \leq \varepsilon\rho, \text{ where } \varepsilon = \sqrt{1-\lambda_{min}}. \quad (4.9)$$

The distance between the approximate $c^{(r)}$ and exact $c$ solutions decreases not slower than infinitely decreasing geometric progression [33]

$$\rho(c^{(r)},c) \leq \varepsilon^r \rho(c^{(0)},c) \leq \frac{\varepsilon^r}{1-\varepsilon} \rho(c^{(0)},c^{(1)}). \quad (4.10)$$

The result obtained implies that the number of iterations required for the distance between the approximate and exact solutions to decrease by the factor of $\exp(k_0)$ is

$$r_0 \approx \frac{-2k_0}{\ln(1-\lambda_{min})}. \quad (4.11)$$

Figure 1 shows an example of the analysis of the iteration procedure convergence. A dependence $\lambda_{min}(\alpha)$ to be found is shown by a solid curve consisted of two segments of parabolas. Besides that, Fig. 1 shows the number of iterations resulted in the same solution for various $\alpha$ at a given accuracy, as well as the approximation of the number of iterations by (4.1) (dotted line). The log likelihood function was controlled in the course of the iteration procedure: the procedure was stopped when the log likelihood function changed by less than $10^{-10}$.

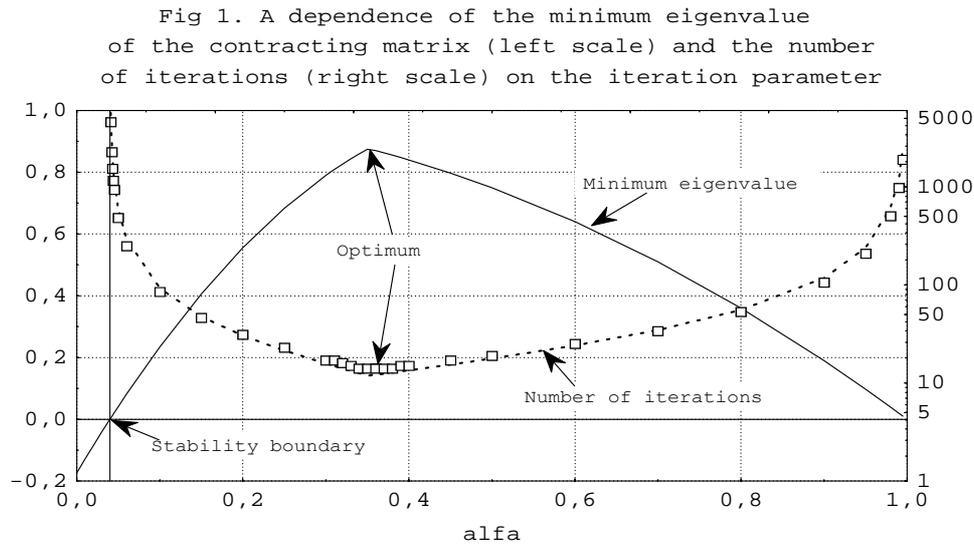

Fig 1. A dependence of the minimum eigenvalue of the contracting matrix (left scale) and the number of iterations (right scale) on the iteration parameter

## 5. Statistical Properties of State Estimator

For the sake of simplicity, consider a real valued psi function.
Let an expansion have the form

$$\psi(x) = \sqrt{1-(c_1^2+...+c_{s-1}^2)}\varphi_0(x) + c_1\varphi_1(x) + ... + c_{s-1}\varphi_{s-1}(x). \quad (5.1)$$

Here, we have eliminated the coefficient $c_0 = \sqrt{1-(c_1^2+...+c_{s-1}^2)}$ from the set of parameters to be estimated, since it is expressed via the other coefficients by the normalization condition.

The parameters $c_1, c_2, ..., c_{s-1}$ are independent. We will study their asymptotic behavior using the Fisher information matrix [22, 26-28]

$$I_{ij}(c) = n \cdot \int \frac{\partial \ln p(x,c)}{\partial c_i} \frac{\partial \ln p(x,c)}{\partial c_j} p(x,c) dx. \quad (5.2)$$

It is of particular importance for our study that the Fisher information matrix drastically simplifies if the psi function is introduced:

$$I_{ij} = 4n \cdot \int \frac{\partial \psi(x,c)}{\partial c_i} \frac{\partial \psi(x,c)}{\partial c_j} dx. \quad (5.3)$$

In the case of the expansion (5.1), the information matrix $I_{ij}$ is $(s-1) \times (s-1)$ matrix of the form

$$I_{ij} = 4n\left(\delta_{ij} + \frac{c_i c_j}{c_0^2}\right), \quad c_0 = \sqrt{1-(c_1^2 + ... + c_{s-1}^2)}. \quad (5.4)$$

A noticeable feature of the expression (5.4) is its independence on the choice of basis functions. Let us show that only the root density estimator has this property.

Consider the following problem that can be referred to as a generalized orthogonal series density estimator. Let the density $p$ be estimated by a composite function of another (for simplicity, real-valued) function $g$. The latter function, in its turn, is represented in the form of the expansion in terms of a set of orthonormal functions, i.e.,

$$p = p(g), \text{ where } g(x) = \sum_{i=0}^{s-1} c_i \varphi_i(x). \quad (5.5)$$

Let the coefficients $c_i;\ i = 0, 1, ..., s-1$ be estimated by the maximum likelihood method.

Consider the following matrix:

$$\tilde{I}_{ij}(c) = n \cdot \int \frac{\partial \ln p(x,c)}{\partial c_i} \frac{\partial \ln p(x,c)}{\partial c_j} p(x,c) dx =$$
$$= n\int \frac{1}{p} \frac{\partial p(x,c)}{\partial c_i} \frac{\partial p(x,c)}{\partial c_j} dx = n\int \frac{1}{p}\left(\frac{\partial p}{\partial g}\right)^2 \frac{\partial g(x,c)}{\partial c_i} \frac{\partial g(x,c)}{\partial c_j} dx \quad (5.6)$$

The structure of this matrix is simplest if its elements are independent of both the density and basis functions. This can be achieved if (and only if) $p(g)$ satisfies the condition

$$\frac{1}{p}\left(\frac{\partial p}{\partial g}\right)^2 = const$$

yielding

$$g = const\sqrt{p}. \quad (5.7)$$

Choosing unity as the constant in the last expression, we arrive at the psi function $g = \psi = \sqrt{p}$ with the simplest normalization condition (2.3).

The $\tilde{I}$ matrix has the form

$$\tilde{I}_{ij} = 4n\delta_{ij} \qquad i,j = 0,1,...,s-1. \quad (5.8)$$

The $\tilde{I}$ matrix under consideration is not the true Fisher information matrix, since the expansion parameters $c_i$ are dependent. They are related to each other by the normalization condition. That is why we will refer to this matrix as a prototype of the Fisher information matrix.

As is seen from the asymptotic expansion of the log likelihood function in the vicinity of a stationary point, statistical properties of the distribution parameters are determined by the quadratic form $\sum_{i,j=0}^{s-1} \tilde{I}_{ij} \delta c_i \delta c_j$. Separating out zero component of the variation and taking into account that

$\delta c_0^2 = \sum_{i,j=1}^{s-1} \frac{c_i c_j}{c_0^2} \delta c_i \delta c_j$ (see the expression (5.12) below), we find

$$\sum_{i,j=0}^{s-1} \tilde{I}_{ij} \delta c_i \delta c_j = \sum_{i,j=1}^{s-1} I_{ij} \delta c_i \delta c_j, \quad (5.9)$$

where the true information matrix $I$ has the form of (5.4).

Thus, the representation of the density in the form $p = \psi^2$ (and only this representation) results in a universal (and simplest) structure of the Fisher information matrix.

In view of the asymptotic efficiency, the covariance matrix of the state estimator is the inverse Fisher information matrix:

$$\Sigma(\hat{c}) = I^{-1}(c) \quad (5.10)$$

The matrix components are

$$\Sigma_{ij} = \frac{1}{4n} (\delta_{ij} - c_i c_j) \quad i, j = 1, \ldots, s-1. \quad (5.11)$$

Now, let us extend the covariance matrix found by appending the covariance between the $c_0$ component of the state vector and the other components.

Note that

$$\delta c_0 = \frac{\partial c_0}{\partial c_i} \delta c_i = \frac{-c_i}{c_0} \delta c_i. \quad (5.12)$$

This yields

$$\Sigma_{0j} = \overline{\delta c_0 \delta c_j} = \frac{-c_i}{c_0} \overline{\delta c_i \delta c_j} =$$

$$= \frac{-\Sigma_{ji} c_i}{c_0} = \frac{-c_i (\delta_{ji} - c_j c_i)}{4 n c_0} = -\frac{c_0 c_j}{4n}. \quad (5.13)$$

Similarly,

$$\Sigma_{00} = \overline{\delta c_0 \delta c_0} = \frac{c_i c_j}{c_0^2} \overline{\delta c_i \delta c_j} = \frac{c_i c_j}{c_0^2} \Sigma_{ij} = \frac{1 - c_0^2}{4n}. \quad (5.14)$$

Finally, we find that the covariance matrix has the same form as (5.11):

$$\Sigma_{ij} = \frac{1}{4n} (\delta_{ij} - c_i c_j) \quad i, j = 0, 1, \ldots, s-1. \quad (5.15)$$

This result seems to be almost evident, since the zero component is not singled out from the others (or more precisely, it has been singled out to provide the fulfillment of the normalization condition). From the geometrical standpoint, the covariance matrix (5.15) is a second-order tensor.

Moreover, the covariance matrix (up to a constant factor) is a single second-order tensor satisfying the normalization condition.

Indeed, according to the normalization condition,

$$\delta(c_i c_i) = 2 c_i \delta c_i = 0. \quad (5.16)$$

Multiplying the last equation by an arbitrary variation $\delta c_j$ and averaging over the statistical ensemble, we find

$$c_i E(\delta c_i \delta c_j) = \Sigma_{ji} c_i = 0 \quad (5.17)$$

Only two different second-order tensors can be constructed on the basis of the vector $c_i$: $\delta_{ij}$ and $c_i c_j$. In order to provide the fulfillment of (5.17) following from the normalization condition, these tensors have to appear in the matrix only in the combination (5.15).

It is useful to consider another derivation of the covariance matrix. According to the normalization condition, the variations $\delta c_i$ are dependent, since they are related to each other by the linear relationship (5.16). In order to make the analysis symmetric (in particular, to avoid expressing one component via the others as it has been done in (5.1)), one may turn to other variables that will be referred to as principle components.

Consider the following unitary (orthogonal) transformation:

$$U_{ij} \delta c_j = \delta f_i \qquad i, j = 0, 1, \ldots, s-1 \quad (5.18)$$

Let the first (to be more precise, zero) row of the transformation matrix coincide with the state vector: $U_{0j} = c_j$. Then, according to (5.16), the zero variation is identically zero in new coordinates: $\delta f_0 = 0$.

The inverse transformation is

$$U^+_{ij} \delta f_j = \delta c_i \qquad i, j = 0, 1, \ldots, s-1 \quad (5.19)$$

In view of the fact that $\delta f_0 = 0$, the first (more precisely, zero) column of the matrix $U^+$ can be eliminated turning the matrix into the factor loadings matrix $L$. Then

$$\delta c_i = L_{ij} \delta f_j \qquad i = 0, 1, \ldots, s-1; \quad j = 1, \ldots, s-1 \quad (5.20)$$

The relationship found shows that $s$ components of the state-vector variation are expressed through $s-1$ principal components (that are independent Gaussian variables).

In terms of principle components, the Fisher information matrix and covariance matrix are proportional to a unit matrix:

$$I^f_{ij} = 4n \delta_{ij} \qquad i, j = 1, \ldots, s-1 \quad (5.21)$$

$$\Sigma^f_{ij} = \overline{\delta f_i \delta f_j} = \frac{\delta_{ij}}{4n} \qquad i, j = 1, \ldots, s-1 \quad (5.22)$$

The last relationship particularly shows that the principal variation components are independent and have the same variance $\dfrac{1}{4n}$.

The expression for the covariance matrix of the state vector components can be easily found on the basis of (5.22). Indeed,

$$\Sigma_{ij} = \overline{\delta c_i \delta c_j} = L_{ik} L_{js} \overline{\delta f_k \delta f_s} = L_{ik} L_{js} \frac{\delta_{ks}}{4n} = \frac{L_{ik} L_{jk}}{4n} \quad (5.23)$$

In view of the unitarity of the $U^+$ matrix, we have

$$L_{ik} L_{jk} + c_i c_j = \delta_{ij} \quad (5.24)$$

Taking into account two last formulas, we finally find the result presented above:

$$\Sigma_{ij} = \frac{1}{4n}\left(\delta_{ij} - c_i c_j\right) \quad i, j = 0, 1, \ldots, s-1. \quad (5.25)$$

In quantum mechanics, the matrix

$$\rho_{ij} = c_i c_j \quad (5.26)$$

is referred to as a density matrix (of a pure state). Thus,

$$\Sigma = \frac{1}{4n}(E - \rho), \quad (5.27)$$

where $E$ is the $s \times s$ unit matrix.

In the diagonal representation,

$$\Sigma = UDU^\dagger, \quad (5.28)$$

where $U$ and $D$ are unitary (orthogonal) and diagonal matrices, respectively.

As is well known from quantum mechanics and readily seen straightforwardly, the density matrix of a pure state has the only (equal to unity) element in the diagonal representation. Thus, in our case, the diagonal of the $D$ matrix has the only element equal to zero (the corresponding eigenvector is the state vector); whereas the other diagonal elements are equal to $\frac{1}{4n}$ (corresponding eigenvectors and their linear combinations form a subspace that is orthogonal complement to the state vector). The zero element at a principle diagonal indicates that the inverse matrix (namely, the Fisher information matrix of the $s$-th order) does not exist. It is clear since there are only $s-1$ independent parameters in the distribution.

The results on statistical properties of the state vector reconstructed by the maximum likelihood method can be summarized as follows. In contrast to a true state vector, the estimated one involves noise in the form of a random deviation vector located in the space orthogonal to the true state vector. The components of the deviation vector (totally, $s-1$ components) are asymptotically normal independent random variables with the same variance $\frac{1}{4n}$. In the aforementioned $s-1$-dimensional space, the deviation vector has an isotropic distribution, and its squared length is the random variable $\frac{\chi^2_{s-1}}{4n}$, where $\chi^2_{s-1}$ is the random variable with the chi-square distribution of $s-1$ degrees of freedom, i.e.,

$$c_i = \left(c, c^{(0)}\right) \cdot c_i^{(0)} + \xi_i \quad i = 0, 1, \ldots, s-1. \quad (5.29)$$

where $c^{(0)}$ and $c$ are true and estimated state vectors, respectively; $\left(c, c^{(0)}\right) = c_i c_i^{(0)}$, their scalar product; and $\xi_i$, the deviation vector. The deviation vector is orthogonal to the vector $c^{(0)}$ and has the squared length of $\frac{\chi^2_{s-1}}{4n}$ determined by chi-square distribution of $s-1$ degrees of freedom, i.e.,

$$\left(\xi, c^{(0)}\right) = \xi_i c_i^{(0)} = 0 \quad \left(\xi, \xi\right) = \xi_i \xi_i = \frac{\chi^2_{s-1}}{4n} \quad (5.30)$$

Squaring (5.29), in view of (5.30), we have

$$1 - \left(c, c^{(0)}\right)^2 = \frac{\chi^2_{s-1}}{4n}. \quad (5.31)$$

This expression means that the squared scalar product of the true and estimated state vectors is smaller than unity by asymptotically small random variable $\dfrac{\chi^2_{s-1}}{4n}$.

The results found allow one to introduce a new stochastic characteristic, namely, a confidence cone (instead of a standard confidence interval). Let $\vartheta$ be the angle between an unknown true state vector $c^{(0)}$ and that $c$ found by solving the likelihood equation. Then,

$$\sin^2 \vartheta = 1 - \cos^2 \vartheta = 1 - (c, c^{(0)})^2 = \frac{\chi^2_{s-1}}{4n} \leq \frac{\chi^2_{s-1,\alpha}}{4n}. \quad (5.32)$$

Here, $\chi^2_{s-1,\alpha}$ is the quantile corresponding to the significance level $\alpha$ for the chi-square distribution of $s-1$ degrees of freedom.

The set of directions determined by the inequality (5.32) constitutes the confidence cone. The axis of a confidence cone is the reconstructed state vector $c$. The confidence cone covers the direction of an unknown state vector at a given confidence level $P = 1 - \alpha$.

From the standpoint of theory of unitary transformations in quantum mechanics (in our case transformations are reduced to orthogonal), it can be found an expansion basis in (5.1) such that the sum will contain the only nonzero term, namely, the true psi function. This result means that if the best basis is guessed absolutely right and the true state vector is $(1,0,0,\ldots,0)$, the empirical state vector estimated by the maximum likelihood method will be the random vector $(c_0, c_1, c_2, \ldots, c_{s-1})$, where $c_0 = \sqrt{1 - (c_1^2 + \ldots + c_{s-1}^2)}$, and the other components $c_i \sim N\left(0, \dfrac{1}{4n}\right)$ $i = 1, \ldots, s-1$ will be independent as it has been noted earlier.

**6. Chi-Square Criterion. Test of the Hypothesis That the Estimated State Vector Equals to the State Vector of a General Population. Estimation of the Statistical Significance of Differences between Two Samples.**

Rewrite (5.31) in the form

$$4n\left(1 - (c, c^{(0)})^2\right) = \chi^2_{s-1}. \quad (6.1)$$

This relationship is a chi-square criterion to test the hypothesis that the state vector estimated by the maximum likelihood method $c$ equals to the state vector of general population $c^{(0)}$.

In view of the fact that for $n \to \infty$ $1 + (c, c^{(0)}) \to 2$, the last inequality may be rewritten in another asymptotically equivalent form

$$4n \sum_{i=0}^{s-1} (c_i - c_i^{(0)})^2 = \chi^2_{s-1}. \quad (6.2)$$

Here, we have taken into account that

$$\sum_{i=0}^{s-1} (c_i - c_i^{(0)})^2 = \sum_{i=0}^{s-1} \left(c_i^2 - 2 c_i c_i^{(0)} + c_i^{(0)2}\right) = \sum_{i=0}^{s-1} 2\left(1 - c_i c_i^{(0)}\right).$$

As is easily seen, the approach under consideration involves the standard chi-square criterion as a particular case corresponding to a histogram basis. Indeed, the chi-square parameter is usually defined as [22, 28]

$$\chi^2 = \sum_{i=0}^{s-1} \frac{\left(n_i - n_i^{(0)}\right)^2}{n_i^{(0)}},$$

where $n_i^{(0)}$ is the number of points expected in $i$-th interval according to theoretical distribution.

In the histogram basis, $c_i = \sqrt{\frac{n_i}{n}}$ is the empirical state vector and $c_i^{(0)} = \sqrt{\frac{n_i^{(0)}}{n}}$, theoretical state vector. Then,

$$\chi^2 = n \sum_{i=0}^{s-1} \frac{\left(c_i^2 - c_i^{(0)2}\right)^2}{c_i^{(0)2}} \to 4n \sum_{i=0}^{s-1} \left(c_i - c_i^{(0)}\right)^2. \quad (6.3)$$

Here, the sign of passage to the limit means that random variables appearing in both sides of (6.3) have the same distribution. We have used also the asymptotic approximation

$$c_i^2 - c_i^{(0)2} = \left(c_i + c_i^{(0)}\right)\left(c_i - c_i^{(0)}\right) \to 2c_i^{(0)}\left(c_i - c_i^{(0)}\right). \quad (6.4)$$

Comparing (6.2) and (6.3) shows that the parameter $\chi^2$ is a random variable with $\chi^2$-distribution of $s-1$ degrees of freedom (if the tested hypothesis is valid). Thus, the standard chi-square criterion is a particular case (corresponding to a histogram basis) of the general approach developed here (that can be used in arbitrary basis).

The chi-square criterion can be applied to test the homogeneity of two different set of observations (samples). In this case, the hypothesis that the observations belong to the same statistical ensemble (the same general population) is tested.

In the case under consideration, the standard chi-square criterion [22, 28] may be represented in new terms as follows:

$$\chi^2 = n_1 n_2 \sum_{i=0}^{s-1} \frac{\left(\frac{n_i^{(1)}}{n_1} - \frac{n_i^{(2)}}{n_2}\right)^2}{n_i^{(1)} + n_i^{(2)}} \to 4\frac{n_1 n_2}{n_1 + n_2} \sum_{i=0}^{s-1} \left(c_i^{(1)} - c_i^{(2)}\right)^2, \quad (6.5)$$

where $n_1$ and $n_2$ are the sizes of the first and second samples, respectively; $n_i^{(1)}$ and $n_i^{(2)}$, the numbers of points in the $i$-th interval; and $c_i^{(1)}$ and $c_i^{(2)}$, the empirical state vectors of the samples. In the left side of (6.5), the chi-square criterion in a histogram basis is presented; in the right side, the same criterion in general case. The parameter $\chi^2$ defined in such a way is a random variable with the $\chi^2$ distribution of $s-1$ degrees of freedom (the sample homogeneity is assumed).

## 7. Optimization of the Number of Harmonics

Let an exact (usually, unknown) psi function be

$$\psi(x) = \sum_{i=0}^{\infty} c_i \varphi_i(x). \quad (7.1)$$

Represent the psi-function estimator in the form

$$\hat{\psi}(x) = \sum_{i=0}^{s-1} \hat{c}_i \varphi_i(x). \quad (7.2)$$

Here, the statistical estimators are denoted by caps in order to distinguish them from exact quantities.

Comparison of two formulas shows that difference between the exact and estimated psi functions is caused by two reasons [34]. First, we neglect $s$-th and higher harmonics by truncating the infinite series. Second, the estimated Fourier series coefficients (with caps) differ from unknown exact values.

Let

$$\hat{c}_i = c_i + \delta c_i. \quad (7.3)$$

Then, in view of the basis orthonormality, the squared deviation of the exact function from the approximate one may be written as

$$F(s) = \int (\psi - \hat{\psi})^2 dx = \sum_{i=0}^{s-1} \delta c_i^2 + \sum_{i=s}^{\infty} c_i^2. \quad (7.4)$$

Introduce the notation

$$Q(s) = \sum_{i=s}^{\infty} c_i^2. \quad (7.5)$$

By implication, $Q(s)$ is deterministic (not a random) variable. As for the first term, we will consider it as a random variable asymptotically related to the chi-square distribution:

$$\sum_{i=0}^{s-1} \delta c_i^2 \sim \frac{\chi_{s-1}^2}{4n},$$

where $\chi_{s-1}^2$ is the random variable with the chi-square distribution of $s-1$ degrees of freedom.

Thus, we find that $F(s)$ is a random variable of the form

$$F(s) = \frac{\chi_{s-1}^2}{4n} + Q(s). \quad (7.6)$$

We will look for an optimal number of harmonics using the condition for minimum of the function mean value $\overline{F}(s)$:

$$\overline{F}(s) = \frac{s-1}{4n} + Q(s) \to \min. \quad (7.7)$$

Assume that, at sufficiently large $s$,

$$Q(s) = \frac{f}{s^r}. \quad (7.8)$$

The optimal value resulting from the condition $\dfrac{\partial \overline{F}(s)}{\partial s} = 0$ is

$$s_{opt} = \sqrt[r+1]{4rfn}. \quad (7.9)$$

The formula (7.9) has a simple meaning: the Fourier series should be truncated when its coefficients become equal to or smaller than the error, i.e., $c_s^2 \leq \dfrac{1}{4n}$. From (7.8) it follows that $c_s^2 \approx \dfrac{rf}{s^{r+1}}$. The combination of the last two formulas yields the estimation (7.9).

The coefficients $f$ and $r$ can be calculated by the regression method. The regression function (7.8) is smoothed by taking a logarithm

$$\ln Q(s) = \ln f - r \ln s \quad (7.10)$$

Another way to numerically minimize $F(s)$ is to detect the step when the inequality $c_s^2 \leq \dfrac{1}{4n}$ is systematically satisfied. It is necessary to determine that the inequality is met just systematically in contrast to the case when several coefficients are equal to zero in result of the symmetry of the function.

Our approach to estimate the number of expansion terms does not pretend to high rigor. For instance, strictly speaking, our estimation of the statistical noise level does not directly concern the coefficients in the infinite Fourier series. Indeed, the estimation was performed in the case when the estimated function is exactly described by a finite (preset) number of terms in the Fourier series with coefficients involving certain statistical error due to the finite size of a sample. Moreover, since the function to be determined is unknown (otherwise, there is no a problem), any estimation of the truncation error is approximate, since it is related to the introduction of additional assumptions.

The optimization of the number of terms in the Fourier series may be performed by the Tikhonov regularization methods [1, 34].

## 8. Numerical Simulations. Chebyshev-Hermite Basis

In this paper, the set of the Chebyshev - Hermite functions corresponding to the stationary states of a quantum harmonic oscillator is used for numerical simulations. In particular, this basis is convenient since the Gaussian distribution in zero-order approximation can be achieved by choosing the ground oscillator state; and adding the contributions of higher harmonics into the state vector provides deviations from the gaussianity.

The set of the Chebyshev-Hermite basis functions is [29, 31]

$$\varphi_k(x) = \frac{1}{\left(2^k k! \sqrt{\pi}\right)^{1/2}} H_k(x) \exp\left(-\frac{x^2}{2}\right), \quad k = 0, 1, 2, \ldots \quad (8.1)$$

Here, $H_k(x)$ is the Chebyshev-Hermite polynomial of the $k$-th order. The first two polynomials have the form

$$H_0(x) = 1, \quad (8.2)$$

$$H_1(x) = 2x. \quad (8.3)$$

The other polynomials can be found by the following recurrent relationship:

$$H_{k+1}(x) - 2x H_k(x) + 2k H_{k-1}(x) = 0. \quad (8.4)$$

The algorithms proposed here have been tested by the Monte Carlo method using the Chebyshev-Hermite functions for a wide range of distributions (mixture of several Gaussian components, gamma distribution, beta distribution etc.). The results of numerical simulations show that the estimation of the number of terms in the Fourier series is close to optimal. It turns out that the approximation accuracy decays more sharply in the case when less terms than optimal are taken into account than in the opposite case when several extra noise harmonics are allowed for. From the aforesaid, it follows that choosing larger number of terms does not result in sharp deterioration of the approximation results. For example, the approximate density of the mixture of two components weakly varies in the range from 8—10 to 50 and more terms for a sample of several hundreds points.

Figure 2 shows an example of comparison of a double-humped curve (dotted line) with an exact distribution (solid line), as well as smoothing the dependence by (7.10) (the sample size is $n = 200$). Note that Figs. 1 and 2 correspond to the same statistical data.

This approach implies that the basis for the psi-function expansion can be arbitrary but it should be preset. In this case, the found results turn out to be universal (independent of basis). This concerns the Fisher information matrix, covariance matrix, chi-square parameter etc. The set of the Chebyshev-Hermite functions can certainly be generalized by introducing translation and scaling parameters that have to be estimated by the maximum likelihood method. The Fisher information matrix, covariance matrix etc. found in such a way would be related only to the Chebyshev-Hermite basis and nothing else.

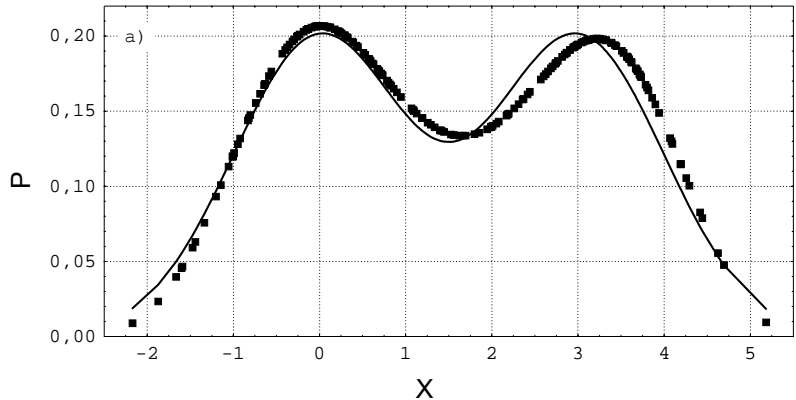

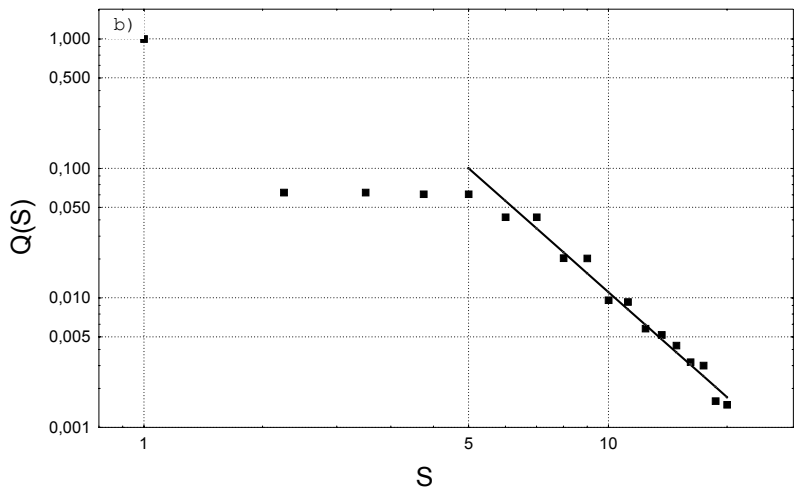

Fig. 2 (a) An example of comparison of a double-humped curve (dotted line) with an exact distribution (solid line); (b) smoothing the dependence by (7.10)

Practically, the expansion basis can be fixed beforehand if the data describe a well-known physical system (e.g., in atomic systems, the basis is preset by nature in the form of the set of stationary states).

In other cases, the basis has to be chosen in view of the data under consideration. For instance, in the case of the Chebyshev-Hermite functions, it can be easily done if one assumes that the distribution is Gaussian in the zero-order approximation.

Note that the formalism presented here is equally applicable to both one-dimensional and multidimensional data. In the latter case, if the Chebyshev-Hermite functions are used, one may assume that multidimensional normal distribution takes place in the zero-order approximation that, in its turn, can be transformed to the standard form by translation, scale, and rotational transformations.

## 9. Density Matrix

The density matrix method is a general quantum mechanical method to study inhomogeneous statistical populations (mixtures). The corresponding technique can be used in statistical data analysis as well.

First, for example, consider a case when the analysis of Sec. 6 shows that two statistical samples are inhomogeneous. In this case, the density matrix represented by a mixture of two components can be constructed for the total population:

$$\rho = \frac{n_1}{n}\rho^{(1)} + \frac{n_2}{n}\rho^{(2)}, \quad (9.1)$$

where

$$n = n_1 + n_2,$$

$$\rho_{ij}^{(1)} = c_i^{(1)} c_j^{(1)*}, \quad (9.2)$$

$$\rho_{ij}^{(2)} = c_i^{(2)} c_j^{(2)*}. \quad (9.3)$$

Any density matrix can be transformed to the diagonal form by a unitary transformation. In the diagonal representation, the density matrix of the pure state will have the only element equal to unity and the other, equal to zero. In the case of two component mixture (9.1), there will be two nonzero elements etc.

Note that only the density matrix of a pure state satisfies the condition

$$\rho^2 = \rho. \quad (9.4)$$

In the diagonal representation, the density for the mixture of $m$ components can be represented in terms of eigenvalues and eigenfunctions of the density matrix:

$$p(x) = \sum_{i=1}^{m} \lambda_i |\psi_i(x)|^2. \quad (9.5)$$

In the case when the first component prevails in the expansion (9.5), it may be considered as responsible for the basic density; whereas the other, describing noise.

Now, we cite some information on the density matrix from quantum mechanics.

The mean value of a physical quantity $A$ is expressed in terms of psi function as follows:

$$\overline{A} = \int \psi^*(x) A(x) \psi(x) dx =$$
$$= c_i c_j^* \int \varphi_j^*(x) A(x) \varphi_i(x) dx = \rho_{ij} A_{ji} = Tr(\rho A). \quad (9.6)$$

Here, the density matrix $\rho$ and the matrix element of $A$ are given by

$$\rho_{ij} = c_i c_j^*, \quad (9.7)$$

$$A_{ji} = \int \varphi_j^*(x) A(x) \varphi_i(x) dx, \quad (9.8)$$

and $Tr$ denotes the trace of a matrix.

The density matrix introduced in such a way relates to a so-called pure state described by a psi function. In general case, a matrix satisfying the following three conditions can be referred to as a density matrix:

1. Hermitian matrix:

$$\rho^+ = \rho. \quad (9.9)$$

Recall that $\rho_{ij}^+ = \rho_{ji}^*$.

2. Positive (nonnegative) matrix

$$\langle z|\rho|z\rangle \equiv \sum_{i,j} \rho_{ij} z_i^* z_j \geq 0 \quad (9.9)$$

for any column vector $|z\rangle$. The sign of equality takes place only for the identically zero column vector. Thus the diagonal elements of the density matrix are always nonnegative.

3. The matrix trace is equal to unity:

$$Tr(\rho) = 1. \quad (9.11)$$

Each density matrix $\rho$ defined on a orthonormal basis $\varphi_i(x)$ may be placed in correspondence with a density operator

$$\rho(x, x_1) = \rho_{ij} \varphi_i^*(x_1) \varphi_j(x). \quad (9.12)$$

In the case when the arguments of the density operator coincide, we obtain the basic object of probability theory, namely, probability density

$$p(x) = \rho(x, x) = \rho_{ij} \varphi_i^*(x) \varphi_j(x). \quad (9.13)$$

The only probability density corresponds to the density matrix (in a given basis). The opposite statement is incorrect.

The mean value (mathematical expectation) of any physical quantity given by an arbitrary operator $A$ is

$$\overline{A} = \rho_{ij} A_{ji} = Tr(\rho A). \quad (9.14)$$

Now, consider the case when the analysis by the algorithm of Sec. 6 shows the homogeneity of both statistical samples. In this case, the state estimator for the total population should be represented by a certain superposition of the state estimators for separate samples. Let us show that the corresponding optimal estimator is the first principle component of the joint density matrix (9.1) in the expansion (9.5).

Let us express the eigenvectors of the joint density matrix (9.1) in terms of eigenvectors of the components:

$$\rho_{ij} c_j = \lambda c_i, \quad (9.15)$$

$$c_j = a_1 c_j^{(1)} + a_2 c_j^{(2)}. \quad (9.16)$$

Substituting (9.16) into (9.15), in view of (9.2) and (9.3), we have a set of the homogeneous equations in two unknowns:

$$\left. \begin{array}{l} (n_1 - \lambda(n_1 + n_2))a_1 + n_1 r a_2 = 0 \\ n_2 r^* a_1 + (n_2 - \lambda(n_1 + n_2))a_2 = 0 \end{array} \right\}, \quad (9.17)$$

where

$$r = (c^{*(1)}, c^{(2)}) = c_i^{*(1)} c_i^{(2)} \quad (9.18)$$

is a scalar product of two vectors.

The system admits a solution if its determinant is equal to zero. Finally, the eigenvalues of the joint density matrix are

$$\lambda_{1,2} = \frac{1 \pm \sqrt{1 - 4k}}{2}, \quad (9.19)$$

where

$$k = \frac{n_1 n_2 (1 - |r|^2)}{(n_1 + n_2)^2}. \quad (9.20)$$

For simplicity's sake, we restrict our consideration to real valued vectors (just as in Sec. 6). In the case of homogeneous samples, asymptotically $1 - r^2 = (1+r)(1-r) \to 2(1-r)$. Then, according to (6.5), we asymptotically have

$$4k = \frac{\chi^2_{s-1}}{n_1 + n_2} \sim O\left(\frac{1}{n_1 + n_2}\right). \quad (9.21)$$

Then,

$$\lambda_1 = 1 - O\left(\frac{1}{n_1 + n_2}\right), \quad (9.22)$$

$$\lambda_2 = O\left(\frac{1}{n_1 + n_2}\right). \quad (9.23)$$

Thus, the first principal component has maximum weight while merging two homogeneous samples. The second component has a weight of an order of $\frac{1}{n_1 + n_2}$ and should be interpreted as a statistical fluctuation. If one drops the second component, the density matrix would become pure.

Equation (9.17) and the normalization condition yield

$$a_1 \approx \frac{n_1}{n_1 + n_2} \quad a_2 \approx \frac{n_2}{n_1 + n_2} \quad (9.24)$$

up to terms of an order of $\frac{1}{n_1 + n_2}$. In view of (5.25), the deviation of the resulting state vector is

$$\xi = a_1 \xi^{(1)} + a_2 \xi^{(2)}, \quad (9.25)$$

$$E(\xi_i \xi_j) = a_1^2 E(\xi_i^{(1)} \xi_j^{(1)}) + a_2^2 E(\xi_i^{(2)} \xi_j^{(2)}) =$$

$$= \frac{1}{4(n_1 + n_2)}(\delta_{ij} - c_i c_j) \quad i, j = 0, 1, ..., s-1. \quad (9.26)$$

The last equation shows that the fist principal component of a joint density matrix is asymptotically efficient estimator of an unknown state vector.

Thus, in merging two large homogeneous samples with certain state estimators, it is not necessary to return to the initial data and solve the likelihood equation for a total population. It is sufficient to find the first principle component of a joint density matrix. This component will be an estimator for the state vector of statistical ensemble that is refined over the sample population. As it has been shown above, such an estimator is asymptotically effective and is not worth than the estimator on the basis of initial data (to be precise, the loss in accuracy is of a higher order of magnitude than (9.26)).

This property is of essential importance. It implies that the estimated psi function involves practically all useful information contained in a large sample. In other words, psi function is asymptotically sufficient statistics. This property can also be interpreted as an asymptotic quasilinearity of a state vector satisfying the nonlinear likelihood equation.

## 10. Phase Role. Statistical Analysis of Mutually Complementing Experiments. Inverse Statistical Problem in Quantum Mechanics.

We have defined the psi function as a complex-valued function with the squared absolute value equal to the probability density. From this point of view, any psi function can be determined up to arbitrary phase factor $\exp(iS(x))$. In particular, the psi function can be chosen real-valued. For instance, in estimating the psi function in a histogram basis, the phases of amplitudes (3.2), which have been chosen equal to zero, could be arbitrary.

At the same time, from the physical standpoint, the phase of psi function is not redundant. The psi function becomes essentially complex valued function in analysis of mutually complementing (according to Bohr) experiments with micro objects [35].

According to quantum mechanics, experimental study of statistical ensemble in coordinate space is incomplete and has to be completed by study of the same ensemble in another (canonically conjugate, namely, momentum) space. Note that measurements of ensemble parameters in canonically conjugate spaces (e.g., coordinate and momentum spaces) cannot be realized in the same experimental setup.

The uncertainty relation implies that the two-dimensional density in phase space $P(x,p)$ is physically senseless, since the coordinates and momenta of micro objects cannot be measured simultaneously. The coordinate $P(x)$ and momentum $\widetilde{P}(p)$ distributions should be studied separately in mutually complementing experiments and then combined by introducing the psi function.

The coordinate-space and momentum-space psi functions are related to each other by the Fourier transform

$$\psi(x) = \frac{1}{\sqrt{2\pi}} \int \widetilde{\psi}(p) \exp(ipx) dp, \quad (10.1)$$

$$\widetilde{\psi}(p) = \frac{1}{\sqrt{2\pi}} \int \psi(x) \exp(-ipx) dx. \quad (10.2)$$

Consider a problem of estimating an unknown psi function ($\psi(x)$ or $\widetilde{\psi}(p)$) by experimental data observed both in coordinate and momentum spaces. We will refer to this problem as an inverse statistical problem of quantum mechanics (do not confuse it with an inverse problem in the scattering theory). The predictions of quantum mechanics are considered as a direct problem. Thus, we consider quantum mechanics as a stochastic theory, i.e., a theory describing statistical (frequency) properties of experiments with random events. However, quantum mechanics is a special stochastic theory, since one has to perform mutually complementing experiments (space-time description has to be completed by momentum-energy one) to get statistically full description of a population (ensemble). In order for various representations to be mutually consistent, the theory should be expressed in terms of probability amplitude rather than probabilities themselves.

A simplified approach to the inverse statistical problem, which will be exemplified by numerical example, may be as follows. Assume that density estimators $P(x)$ and $\widetilde{P}(p)$ have already been found (e.g., by histogram estimation). It is required to approximate the psi function for a statistical ensemble. Figure 3 shows the comparison between exact densities that could be calculated if the psi function of an ensemble is known (solid line), and histogram estimators obtained in mutually complementing experiments. In each experiment, the sample size is 10000 points. In Fig. 4, the exact psi function is compared to that estimated by samples. The solution was found by iteration procedure of adjusting the phase of psi function in coordinate and momentum representations. In zero-order approximation ($(r=0)$), the phases were assumed to be zero. The momentum-space phase in the $r+1$ approximation was determined by the Fourier transform of the psi function in the $r$ approximation in the coordinate space and vice versa.

The histogram density estimator results in the discretization of distributions, and hence, natural use of the discrete Fourier transform instead of a continuous one.

From (10.1) and (10.2), we straightforwardly have

$$\int \frac{\partial \psi^*(x)}{\partial x} \frac{\partial \psi(x)}{\partial x} dx = \int p^2 \widetilde{\psi}^*(p) \widetilde{\psi}(p) dp. \quad (10.3)$$

From the standpoint of quantum mechanics, the formula (10.3) implies that the same quantity, namely, the mean square momentum, is defined in two different representations (coordinate and momentum). This quantity has a simple form in the momentum representation, whereas in the coordinate representation, it is rather complex characteristic of distribution shape (irregularity). The corresponding quantity is proportional to the Fisher information on the translation parameter of the distribution center (see Sec. 12).

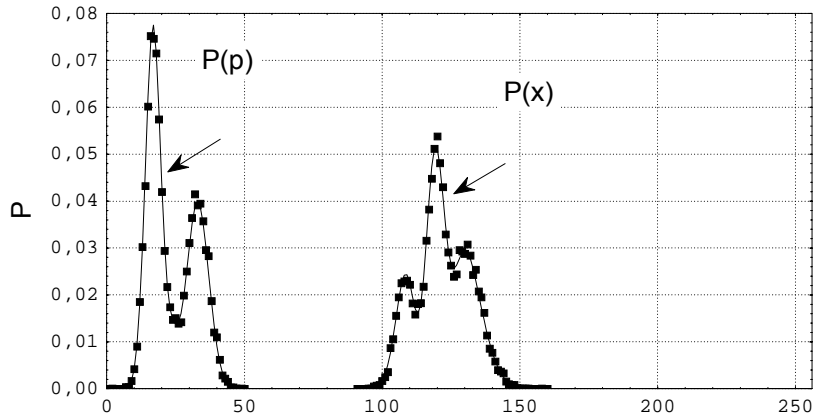

Fig 3. Comparison between exact densities (solid lines) and histogram estimators (dots) in coordinate and momentum spaces.

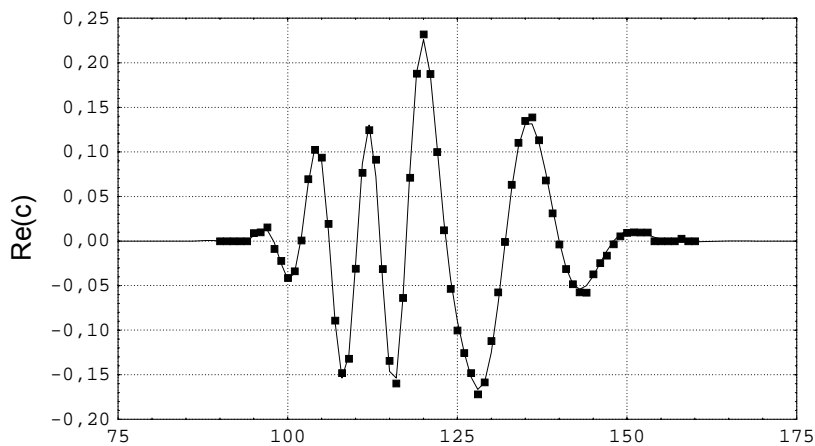

Fig. 4 Comparison between exact psi function (solid line) and that estimated by a sample (dots).

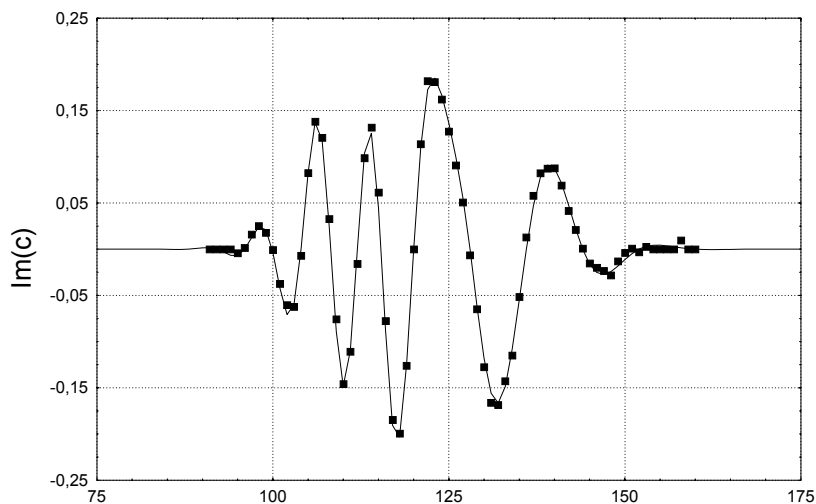

The irregularity cannot be measured in the coordinate space in principle, since it refers to another (momentum) space. In other words, if there is no any information from the canonically conjugate space, the distribution irregularity in the initial space may turn out to be arbitrary high. Singular distributions used in probability theory can serve as density models with an infinite irregularity. From mathematical statistics, it is well-known that arbitrary large sample size does not allow one to determine whether the distribution under consideration is continuous or singular. This causes the ill-posedness of the inverse problem of the probability theory that has already been discussed in Introduction.

Thus, from the standpoint of quantum mechanics, the ill-posedness of the classical problem of density estimation by a sample is due to lack of information from the canonically conjugate space. Regularization methods for inverse problem consist in excluding a priori strongly-irregular functions from consideration. This is equivalent to suppression of high momenta in the momentum space.

Let us turn now to more consistent description of the method for estimation of the state vector of a statistical ensemble on the basis of experimental data obtained in mutually complementing experiments. Consider corresponding generalization of the maximum likelihood principle and likelihood equation. To be specific, we will assume that corresponding experiments relate to coordinate and momentum spaces.

We define the likelihood function as (compare to (1.1))

$$L(x, p|c) = \prod_{i=1}^{n} P(x_i|c) \prod_{j=1}^{m} \tilde{P}(p_j|c). \quad (10.4)$$

Here, $P(x_i|c)$ and $\tilde{P}(p_j|c)$ are the densities in mutually complementing experiments corresponding to the same state vector $c$. We assume that $n$ measurements were made in the coordinate space; and $m$, in the momentum one.

Then, the log likelihood function has the form (instead of (1.2))

$$\ln L = \sum_{i=1}^{n} \ln P(x_i|c) + \sum_{j=1}^{m} \ln \tilde{P}(p_j|c). \quad (10.5)$$

The maximum likelihood principle together with the normalization condition evidently results in the problem of maximization of the following functional:

$$S = \ln L - \lambda(c_i c_i^* - 1), \quad (10.6)$$

where $\lambda$ is the Lagrange multiplier and

$$\ln L = \sum_{k=1}^{n} \ln(c_i c_j^* \varphi_i(x_k) \varphi_j^*(x_k)) + \sum_{l=1}^{m} \ln(c_i c_j^* \tilde{\varphi}_i(p_l) \tilde{\varphi}_j^*(p_l)). \quad (10.7)$$

Here, $\tilde{\varphi}_i(p)$ is the Fourier transform of the function $\varphi_i(x)$.

The likelihood equation has the form similar to (2.10)

$$R_{ij} c_j = \lambda c_i \qquad i, j = 0, 1, \ldots, s-1, \quad (10.8)$$

where the $R$ matrix is determined by

$$R_{ij} = \sum_{k=1}^{n} \frac{\varphi_i^*(x_k) \varphi_j(x_k)}{P(x_k)} + \sum_{l=1}^{m} \frac{\tilde{\varphi}_i^*(p_l) \tilde{\varphi}_j(p_l)}{\tilde{P}(p_l)}. \quad (10.9)$$

By full analogy with calculations conducted in Sec. 2, it can be proved that the most likely state vector always corresponds to the eigenvalue $\lambda = n + m$ of the $R$ matrix (equal to sum of measurements).

The likelihood equation can be easily expressed in the form similar to (2.13):

$$\frac{1}{n+m}\left(\sum_{k=1}^{n}\frac{\varphi_i^*(x_k)}{\sum_{j=1}^{s}c_j^*\varphi_j^*(x_k)}+\sum_{l=1}^{m}\frac{\tilde{\varphi}_i^*(p_l)}{\sum_{j=1}^{s}c_j^*\tilde{\varphi}_j^*(p_l)}\right)=c_i. \quad (10.10)$$

The Fisher information matrix (prototype) is determined by the total information contained in mutually complementing experiments (compare to (5.2) and (5.3)):

$$\tilde{I}_{ij}(c)=n\cdot\int\frac{\partial\ln P(x,c)}{\partial c_i}\frac{\partial\ln P(x,c)}{\partial c_j^*}P(x,c)dx+m\cdot\int\frac{\partial\ln\tilde{P}(p,c)}{\partial c_i}\frac{\partial\ln\tilde{P}(p,c)}{\partial c_j^*}\tilde{P}(p,c)dp, \quad (10.11)$$

$$\tilde{I}_{ij}=n\cdot\int\frac{\partial\psi(x,c)}{\partial c_i}\frac{\partial\psi^*(x,c)}{\partial c_j^*}dx+m\cdot\int\frac{\partial\tilde{\psi}(p,c)}{\partial c_i}\frac{\partial\tilde{\psi}^*(p,c)}{\partial c_j^*}dp=(n+m)\delta_{ij}. \quad (10.12)$$

Note that the factor of 4 is absent in (10.12) in contrast to the similar formula (5.3). This is because of the fact that it is necessary to distinguish $\psi(x)$ and $\psi^*(x)$ as well as $c$ and $c^*$.

Consider the following simple transformation of a state vector that is of vital importance (global gauge transformation). It is reduced to multiplying the initial state vector by arbitrary phase factor:

$$c'=\exp(i\alpha)c, \quad (10.13)$$

where $\alpha$ is arbitrary real number.

One can easily verify that the likelihood function is invariant against the gauge transformation (10.13). This implies that the state vector can be estimated by experimental data up to arbitrary phase factor. In other words, two state vectors that differ only in a phase factor describe the same statistical ensemble. The gauge invariance, of course, also manifests itself in theory, e.g., in the gauge invariance of the Schrödinger equation.

The variation of a state vector that corresponds to infinitesimal gauge transformation is evidently

$$\delta c_j=i\alpha\, c_j \qquad j=0,1,\ldots,s-1, \quad (10.14)$$

where $\alpha$ is a small real number.

Consider how the gauge invariance has to be taken into account in considering statistical fluctuations of the components of a state vector. The normalization condition ($c_j c_j^*=1$) yields that the variations of the components of a state vector satisfy the condition

$$c_j\delta c_j^*+(\delta c_j)c_j^*=0. \quad (10.15)$$

Here, $\delta c_j=\hat{c}_j-c_j$ is the deviation of the state estimator found by the maximum likelihood method from the true state vector characterizing the statistical ensemble.

In view of the gauge invariance, let us divide the variation of a state vector into two terms $\delta c=\delta_1 c+\delta_2 c$. The first term $\delta_1 c=i\alpha\, c$ corresponds to gauge arbitrariness, and the second one $\delta_2 c$ is a real physical fluctuation.

An algorithm of dividing of the variation into gauge and physical terms can be represented as follows. Let $\delta c$ be arbitrary variation meeting the normalization condition. Then, (10.15) yields $(\delta c_j)c_j^*=i\varepsilon$, where $\varepsilon$ is a small real number.

Dividing the variation $\delta c$ into two parts in this way, we have

$$(\delta c_j)c_j^*=(i\alpha c_j+\delta_2 c_j)c_j^*=i\alpha+(\delta_2 c_j)c_j^*=i\varepsilon. \quad (10.16)$$

Choosing the phase of the gauge transformation according to the condition $\alpha=\varepsilon$, we find

$$(\delta_2 c_j) c_j^* = 0. \quad (10.17)$$

Let us show that this gauge transformation provides minimization of the sum of squares of variation absolute values. Let $(\delta c_j) c_j^* = i\varepsilon$. Having performed infinitesimal gauge transformation, we get the new variation

$$\delta c_j' = -i\alpha c_j + \delta c_j. \quad (10.18)$$

Our aim is to minimize the following expression:

$$\delta c_j' \delta c_j'^* = (-i\alpha c_j + \delta c_j)(i\alpha c_j^* + \delta c_j^*) = \delta c_j \delta c_j^* - 2\varepsilon\alpha + \alpha^2 \to \min. \quad (10.19)$$

Evidently, the last expression has a minimum at $\alpha = \varepsilon$.

Thus, the gauge transformation providing separation of the physical fluctuation from the variation achieves two aims.

First, the condition (10.15) is divided into two independent conditions:

$$(\delta c_j) c_j^* = 0 \text{ and } (\delta c_j^*) c_j = 0 \quad (10.20).$$

Here, we have dropped the subscript 2 assuming that the state vector variation is a physical fluctuation free of the gauge component).

Second, this transformation results in mean square minimization of possible variations of a state vector.

Let $\delta c$ be a column vector, then the Hermitian conjugate value $\delta c^+$ is a row vector. Statistical properties of the fluctuations are determined by the quadratic form $\delta c^+ \widetilde{I} \delta c = \sum_{i,j=0}^{s-1} \widetilde{I}_{ij} \delta c_j \delta c_i^*$. In order to switch to independent variables, we will explicitly express (as in Sec. 5) a zero component in terms of the others. According to (10.20), we have $\delta c_0 = -\dfrac{c_j^* \delta c_j}{c_0^*}$. This leads us to $\delta c_0 \delta c_0^* = \dfrac{c_i c_j^*}{|c_0|^2} \delta c_j \delta c_i^*$. The quadratic form under consideration can be represented in the form $\sum_{i,j=0}^{s-1} \widetilde{I}_{ij} \delta c_j \delta c_i^* = \sum_{i,j=1}^{s-1} I_{ij} \delta c_j \delta c_i^*$, where the true Fisher information matrix has the form (compare to (5.4))

$$I_{ij} = (n+m)\left(\delta_{ij} + \frac{c_i c_j^*}{|c_0|^2}\right) \quad i,j = 1,...,s-1, \quad (10.21)$$

where

$$|c_0| = \sqrt{1 - (|c_2|^2 + ... + |c_{s-1}|^2)}. \quad (10.22)$$

The inversion of the Fisher matrix yields the truncated covariance matrix (without zero component). Having calculated covariations with zero components in an explicit form, we finally find the expression for the total covariance matrix that is similar to (5.15):

$$\Sigma_{ij} = \overline{\delta c_i \delta c_j^*} = \frac{1}{(n+m)}(\delta_{ij} - c_i c_j^*) \quad i,j = 0,1,...,s-1. \quad (10.23)$$

The Fisher information matrix and covariance matrix are Hermitian. It is easy to see that the covariance matrix (10.23) satisfies the condition similar to (5.17):

$$\Sigma_{ij} c_j = 0. \quad (10.24)$$

By full analogy with the reasoning of Sec. 5, it is readily seen that the matrix (10.23) is the only (up to a factor) Hermitian tensor of the second order that can be constructed from a state vector satisfying the normalization condition.

The formula (10.23) can be evidently written in the form

$$\Sigma = \frac{1}{(n+m)}(E-\rho), \quad (10.25)$$

where $E$ is the $s \times s$ unit matrix, and $\rho$ is the density matrix.

In the diagonal representation

$$\Sigma = UDU^+, \quad (10.26)$$

where $U$ is the unitary matrix, and $D$ is the diagonal matrix.

The diagonal of the $D$ matrix has the only zero element (the corresponding eigenvector is the state vector). The other diagonal elements are equal to $\frac{1}{n+m}$ (the corresponding eigenvectors and their linear combinations form subspace that is orthogonal complement to a state vector).

The chi-square criterion determining whether the scalar product between the estimated and true vectors $cc^{*(0)}$ is close to unity that is similar to (6.1) is

$$(n+m)\left(1-\left|cc^{*(0)}\right|^2\right) = \chi^2_{s-1}. \quad (10.27)$$

This method is illustrated in Fig. 5. In this figure, the density estimator is compared to the true densities in coordinate and momentum spaces.

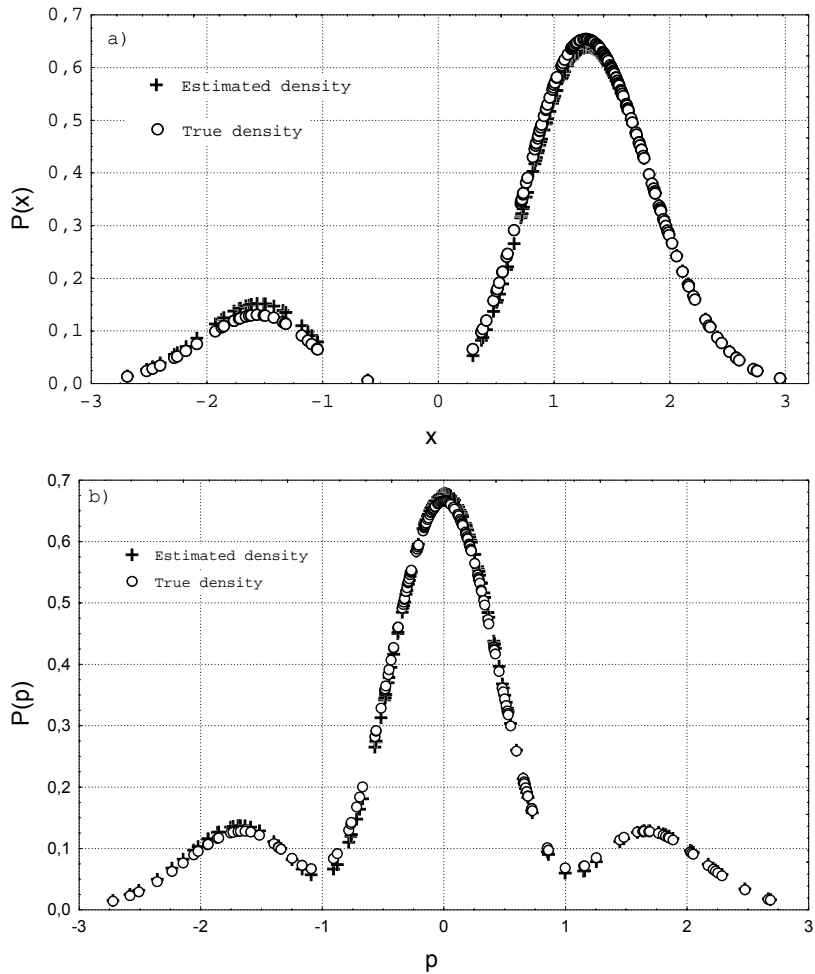

Fig. 5. Comparison between density estimators and true densities in (a) coordinate and (b) momentum spaces

The sample of a size of $n = m = 200$ ($n+m = 400$) was taken from a statistical ensemble of harmonic oscillators with a state vector with three nonzero components ($s = 3$).

## 11. Constraint on the Energy

As it has been already noted, the estimation of a state vector is associated with the problem of suppressing high frequency terms in the Fourier series. In this section, we consider the regularization method based on the law of conservation of energy. Consider an ensemble of harmonic oscillators (although formal expressions are written in general case). Taking into account the normalization condition without any constraints on the energy, the terms may appear that make a negligible contribution to the norm but arbitrary high contribution to the energy. In order to suppress these terms, we propose to introduce both constraints on the norm and energy in the maximum likelihood method. The energy is readily estimated by the data of mutually complementing experiments.

It is worth noting that in the case of potentials with a finite number of discrete levels in quantum mechanics [29, 31], the problem of truncating the series does not arise (if solutions bounded at infinity are considered).

We assume that the psi function is expanded in a series in terms of eigenfunctions of the energy operator $\hat{H}$ (Hamiltonian):

$$\psi(x) = \sum_{i=0}^{s-1} c_i \varphi_i(x), \quad (11.1)$$

where basis functions satisfy the equation

$$\hat{H}\varphi_i(x) = E_i \varphi_i(x). \quad (11.2)$$

Here, $E_i$ is the energy level corresponding to $i$-th state.

The mean energy corresponding to a statistical ensemble with a wave function $\psi(x)$ is

$$\overline{E} = \int \psi^*(x) \hat{H} \psi(x) dx = \sum_{i=0}^{s-1} E_i c_i^* c_i. \quad (11.3)$$

In arbitrary basis

$$\overline{E} = \sum_{i=0}^{s-1} H_{ij} c_j c_i^*, \text{ where } H_{ij} = \int \varphi_i^*(x) \hat{H} \varphi_j(x) dx. \quad (11.4)$$

Consider a problem of finding a maximum likelihood estimator of a state vector in view of a constraint on the energy and norm of the state vector. In energy basis, the problem is reduced to maximization of the following functional:

$$S = \ln L - \lambda_1 (c_i c_i^* - 1) - \lambda_2 (E_i c_i^* c_i - \overline{E}), \quad (11.5)$$

where $\lambda_1$ and $\lambda_2$ are the Lagrange multipliers and $\ln L$ is given by (10.7).
In this case, the likelihood equation has the form

$$R_{ij} c_j = (\lambda_1 + \lambda_2 E_i) c_i, \quad (11.6)$$

where the $R$ matrix is determined by (10.9).

In arbitrary basis, the variational functional and the likelihood equation have the forms

$$S = \ln L - \lambda_1 (c_i c_i^* - 1) - \lambda_2 (H_{ij} c_j c_i^* - \overline{E}), \quad (11.7)$$

$$(R_{ij} - \lambda_2 H_{ij}) c_j = \lambda_1 c_i. \quad (11.8)$$

Having multiplied the both parts of (11.6) (or (11.8)) by $c_i^*$ and summed over $i$, we obtain the same result representing the relationship between the Lagrange multipliers:

$$(n+m) = \lambda_1 + \lambda_2 \overline{E} \quad (11.9)$$

The sample mean energy $\overline{E}$ (i.e., the sum of the mean potential energy that can be measured in the coordinate space and the mean kinetic energy measured in the momentum space) was used as the estimator of the mean energy in numerical simulations below.

Now, let us turn to the study of statistical fluctuations of a state vector in the problem under consideration. We restrict our consideration to the energy representation in the case when the expansion basis is formed by stationary energy states (that are assumed to be nondegenerate).

Additional condition (11.3) related to the conservation of energy results in the following relationship between the components

$$\delta \overline{E} = \sum_{j=0}^{s-1} \left( E_j c_j^* \delta c_j + E_j \delta c_j^* c_j \right) = 0 \quad (11.10)$$

It turns out that both parts of the equality can be reduced to zero independently if one assumes that a state vector to be estimated involves a time uncertainty, i.e., may differ from the true one by a small time translation. The possibility of such a translation is related to the time-energy uncertainty relation.

The well-known expansion of the psi function in terms of stationary energy states, in view of time dependence, has the form ($\hbar = 1$)

$$\psi(x) = \sum_j c_j \exp(-iE_j(t-t_0)) \varphi_j(x) =$$
$$= \sum_j c_j \exp(iE_j t_0) \exp(-iE_j t) \varphi_j(x) \quad (11.11)$$

In the case of estimating the state vector up to translation in time, the transformation

$$c_j' = c_j \exp(iE_j t_0) \quad (11.12)$$

related to arbitrariness of zero-time reference $t_0$ may be used to fit the estimated state vector to the true one.

The corresponding infinitesimal time translation leads us to the following variation of a state vector:

$$\delta c_j = it_0 E_j c_j \quad (11.13)$$

Let $\delta c$ be any variation meeting both the normalization condition and the law of energy conservation. Then, from (10.15) and (11.3) it follows that

$$\sum_j (\delta c_j) c_j^* = i\varepsilon_1 \quad (11.14)$$

$$\sum_j (\delta c_j) E_j c_j^* = i\varepsilon_2 \quad (11.15)$$

where $\varepsilon_1$ and $\varepsilon_2$ are arbitrary small real numbers.

In analogy with Sec. 10, we divide the total variation $\delta c$ into unavoidable physical fluctuation $\delta_2 c$ and variations caused by the gauge and time invariances:

$$\delta c_j = i\alpha c_j + it_0 E_j c_j + \delta_2 c_j \quad (11.16)$$

We will separate out the physical variation $\delta_2 c$, so that it fulfills the conditions (11.14) and (11.15) with a zero right part. It is possible if the transformation parameters $\alpha$ and $t_0$ satisfy the following set of linear equations:

$$\left.\begin{array}{l}\alpha + \overline{E}t_0 = \varepsilon_1 \\ \overline{E}\alpha + \overline{E^2}t_0 = \varepsilon_2\end{array}\right\} \quad (11.17)$$

The determinant of (11.17) is the energy variance

$$\sigma_E^2 = \overline{E^2} - \overline{E}^2 \quad (11.18)$$

We assume that the energy variance is a positive number. Then, there exists a unique solution of the set (11.17). If the energy dissipation is equal to zero, the state vector has the only nonzero component. In this case, the gauge transformation and time translation are dependent, since they are reduced to a simple phase shift.

In full analogy with the reasoning on the gauge invariance, one can show that in view of both the gauge invariance and time homogeneity, the transformation satisfying (11.17) provides minimization of the total variance of the variations (sum of squares of the components absolute values). Thus, one may infer that physical fluctuations are minimum possible fluctuations compatible with the conservation of norm and energy.

Assuming that the total variations are reduced to the physical ones, we assume hereafter that

$$\sum_j (\delta c_j) c_j^* = 0, \quad (11.19)$$

$$\sum_j (\delta c_j) E_j c_j^* = 0. \quad (11.20)$$

The relationships found yield (in analogy with Sec. 10) the conditions for the covariance matrix $\Sigma_{ij} = \overline{\delta c_i \delta c_j^*}$:

$$\sum_j (\Sigma_{ij} c_j) = 0, \quad (11.21)$$

$$\sum_j (\Sigma_{ij} E_j c_j) = 0. \quad (11.22)$$

Consider the unitary matrix $U^+$ with the following two rows (zero and first):

$$(U^+)_{0j} = c_j^*, \quad (11.23)$$

$$(U^+)_{1j} = \frac{(E_j - \overline{E})c_j^*}{\sigma_E} \quad j = 0,1,\ldots,s-1. \quad (11.24)$$

This matrix determines the transition to principle components of the variation

$$U^+_{ij} \delta c_j = \delta f_i. \quad (11.25)$$

According to (11.19) and (11.20), we have $\delta f_0 = \delta f_1 = 0$ identically in new variables so that there remain only $s - 2$ independent degrees of freedom.

The inverse transformation is

$$U\delta f = \delta c \quad (11.26)$$

On account of the fact that $\delta f_0 = \delta f_1 = 0$, one may drop two columns (zero and first) in the $U$ matrix turning it into the factor loadings matrix $L$

$$L_{ij}\delta f_j = \delta c_i \qquad i = 0,1,\ldots,s-1; \quad j = 2,3,\ldots,s-1 \quad (11.27)$$

The $L$ matrix has $s$ rows and $s-2$ columns. Therefore, it provides the transition from $s-2$ independent variation principle components to $s$ components of the initial variation.

In principal components, the Fisher information matrix and covariance matrix are given by

$$I_{ij}^f = (n+m)\delta_{ij}, \quad (11.28)$$

$$\Sigma_{ij}^f = \overline{\delta f_i \delta f_j^*} = \frac{1}{(n+m)}\delta_{ij}. \quad (11.29)$$

In order to find the covariance matrix for the state vector components, we will take into account the fact that the factor loadings matrix $L$ differs form the unitary matrix $U$ by the absence of two aforementioned columns, and hence,

$$L_{ik}L_{kj}^+ = \delta_{ij} - c_i c_j^* - \frac{(E_i - \overline{E})(E_j - \overline{E})}{\sigma_E^2} c_i c_j^*, \quad (11.30)$$

$$\Sigma_{ij} = \overline{\delta c_i \delta c_j^*} = L_{ik}L_{jr}^* \overline{\delta f_k \delta f_r^*} = \frac{L_{ik}L_{kj}^+}{n+m}. \quad (11.31)$$

Finally, the covariance matrix in the energy representation takes the form

$$\Sigma_{ij} = \frac{1}{(n+m)}\left(\delta_{ij} - c_i c_j^*\left(1 + \frac{(E_i - \overline{E})(E_j - \overline{E})}{\sigma_E^2}\right)\right). \quad (11.32)$$

$i, j = 0,1,\ldots,s-1$

It is easily verified that this matrix satisfies the conditions (11.21) and (11.22) resulting from the conservation of norm and energy.

The mean square fluctuation of the psi function is

$$\int \overline{\delta\psi\delta\psi^*}dx = \int \overline{\delta c_i \varphi_i(x)\delta c_j^* \varphi_j^*(x)}dx = \overline{\delta c_i \delta c_i^*} = Tr(\Sigma) = \frac{s-2}{n+m}. \quad (11.33)$$

The estimation of optimal number of harmonics in the Fourier series, similar to that in Sec. 7, has the form

$$s_{opt} = \sqrt[r+1]{rf(n+m)}, \quad (11.34)$$

where the parameters $r$ and $f$ determine the asymptotics for the sum of squares of residuals:

$$Q(s) = \sum_{i=s}^{\infty}|c_i|^2 = \frac{f}{s^r}. \quad (11.35)$$

The norm existence implies only that $r > 0$. In the case of statistical ensemble of harmonic oscillators with existing energy, $r > 1$. If the energy variance is defined as well, $r > 2$.

Figure 6 shows how the constraint on energy decreases high-energy noise. The momentum-space density estimator disregarding the constraint on energy (upper plot) is compared to that accounting for the constraint.

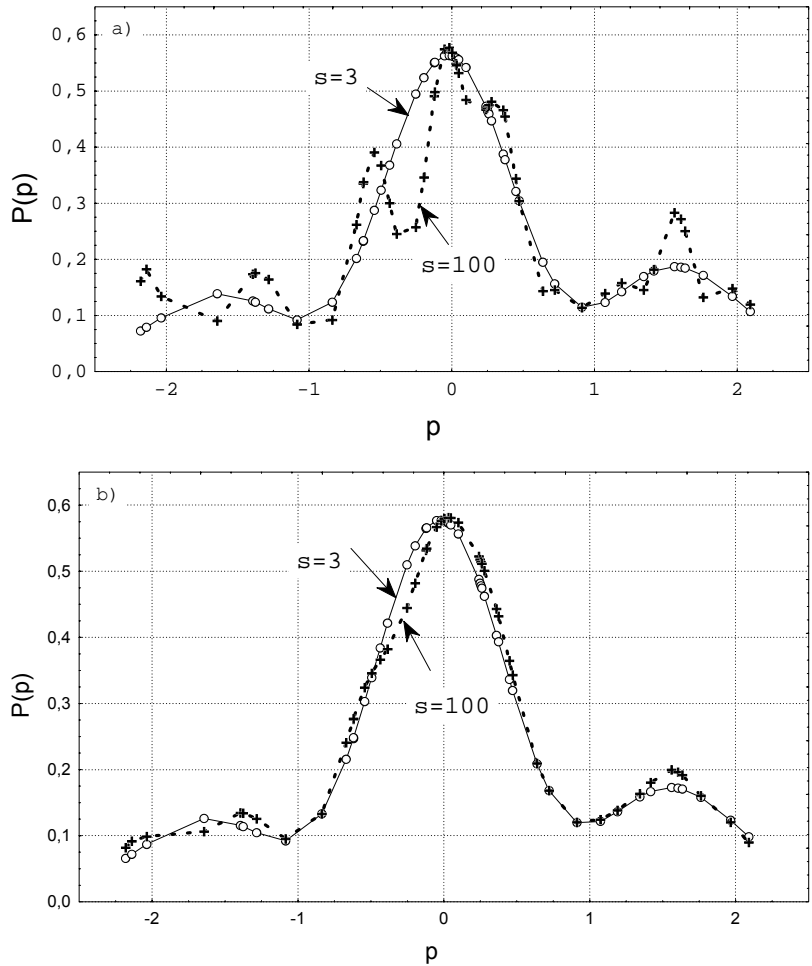

Fig. 6 (a) An estimator without the constraint on energy;
(b) An estimator with the constraint on energy.

The sample of the size of $n = m = 50$ ($n + m = 100$) was taken from the statistical ensemble of harmonic oscillators. The state vector of the ensemble had three nonzero components ($s = 3$). Figure 6 shows the calculation results in bases involving $s = 3$ and $s = 100$ functions, respectively. In the latter case, the number of basis functions coincided with the total sample size. Figure 6 shows that in the case when the constraint on energy was taken into account, the 97 additional noise components influenced the result much weaker than in the case without the constraint.

## 12. Fisher Information and Variational Principle in Quantum Mechanics

The aim of this section is to show a certain relation between the mathematical problem of minimizing the energy by the variational principle in quantum mechanics and the problem of minimizing the Fisher information (more precisely, the Fisher information on the translation parameter) that may be used (and is really used) in some model problems of statistical data analysis (for detail, see [36, 37]).

Let us show that there exists a certain analogy between the Fisher information on the translation parameter and kinetic energy in quantum mechanics. This analogy results in the fact that variational problems of robust statistics are mathematically equivalent to the problems of finding the ground state of a stationary Schrödinger equation [36, 37].

Indeed, kinetic energy in quantum mechanics is (see, e.g., [29]; for simplicity, we consider one-dimensional case):

$$T = -\frac{\hbar^2}{2m}\int \psi^*(x)\frac{\partial^2 \psi(x)}{\partial x^2}dx = \frac{\hbar^2}{2m}\int \frac{\partial \psi^*}{\partial x}\frac{\partial \psi}{\partial x}dx. \quad (12.1)$$

Here, $m$ is the mass of a particle and $\hbar$ is the Planck's constant.

The last equality follows from integration by parts in view of the fact that the psi function and its derivative are equal to zero at the infinity.

Assume that the psi function is real-valued and equals to the square root of the probability density:

$$\psi(x) = \sqrt{p(x)}. \quad (12.2)$$

Then $\frac{\partial \psi}{\partial x} = \frac{p'}{2\sqrt{p}}$, and hence,

$$T = \frac{\hbar^2}{8m}\int \frac{(p')^2}{p}dx = \frac{\hbar^2}{8m}\int \left(\frac{p'}{p}\right)^2 p\,dx = \frac{\hbar^2}{8m}I(p). \quad (12.3)$$

Here, we have taken into account that the Fisher information on the translation parameter, by definition, is [36]

$$I(p) = \int_{-\infty}^{+\infty}\left(\frac{p'(x)}{p}\right)^2 \cdot p(x)\cdot dx. \quad (12.4)$$

The Fisher information (12.4) is a functional with respect to the distribution density $p(x)$.

Let us consider the following variational problem: it is required to find a distribution density $p(x)$ minimizing the Fisher information (12.4) at a given constraints on the mean value of the loss function

$$\overline{U} = \int U(x)p(x)dx \leq U_0. \quad (12.5)$$

In terms of quantum mechanics, the "loss function" $U(x)$ is a potential.

The problem under consideration is linearized if the square root of the density, i.e., psi function is considered instead of the density itself. The variational problem is evidently reduced to minimization of the following functional:

$$S(\psi) = \int_{-\infty}^{+\infty}(\psi'(x))^2 \cdot dx - \lambda_1\left(\int \psi^2(x)dx - 1\right) + \\ + \lambda_2\left(\int U(x)\psi^2(x)dx - U_0\right) \quad (12.6)$$

Here, $\lambda_1$ and $\lambda_2$ are the Lagrange multipliers providing constraints on the norm of a state vector and the loss function, respectively.

From the Lagrange-Euler equation it follows the equation for the psi function

$$-\psi'' + \lambda_2 U\psi = \lambda_1 \psi. \quad (12.7)$$

The last equation turns into a stationary Schrödinger equation if one introduces the notation

$$\frac{\hbar^2}{2m} = \frac{1}{\lambda_2} \qquad E = \frac{\lambda_1}{\lambda_2}. \quad (12.8)$$

Minimization of kinetic energy at a given constraints on potential energy is equivalent to minimization of the total energy. Therefore, the solution of the corresponding problem is the ground state for a particle in a given field. The corresponding result is well-known in quantum mechanics as the variational principle. It is frequently used to estimate the energy of a ground state.

The kinetic energy representations in two different forms (12.1) and (12.3) are known at least from the works by Bohm on quantum mechanics ([38], see also [30]).

The variational principle considered here is employed in papers on robust statistics developed, among others, by Huber [36]. The aim of robust procedures is, first of all, to find such estimators of distribution parameters (e.g., translation parameters) that would be stable against (weakly sensible to) small deviations of a real distribution from the theoretical one. A basic model in this approach is a certain given distribution (usually, Gaussian distribution) with few given outlying observations.

For example, if the estimator of the translation parameter is of the M- type (i.e., estimators of maximum likelihood type), the maximum estimator variance (due to its efficiency) will be determined by minimal Fisher information characterizing the distribution in a given neighborhood [36].

Minimization of the Fisher information shows the way to construct robust distributions. As is seen from the definition (12.4), the Fisher information is a positive quantity making it possible to estimate the complexity of the density curve. Indeed, the Fisher information is related to the squared derivative of the distribution density; therefore, the more complex, irregular, and oscillating the distribution density, the greater the Fisher information. From this point of view, the simplicity can be achieved by providing minimization of the Fisher information at given constraints. The Fisher information may be considered as a penalty function for the irregularity of the density curve. The introduction of such penalty functions aims at regularization of data analysis problems and is based on the compromise between two tendencies: to obtain the data description as detailed as possible using functions without fast local variations [37, 39-41].

In the work by Good and Gaskins [39], the problem of minimization of smoothing functional, which is equal to the difference between the Fisher information and log likelihood function, is stated in order to approximate the distribution density. The corresponding method is referred to as the maximum penalized likelihood method [40-41].

Among all statistical characteristics, the most popular are certainly the sample mean (estimation of the center of probability distribution) and sample variance (to estimate the deviation). Assuming that these are the only parameters of interest, let us find the simplest distribution (in terms of the Fisher information). The corresponding variational problem is evidently equivalent to the problem of finding the minimum energy solution of the Schrödinger equation (12.7) with a quadratic potential. The corresponding solution (density of the ground state of a harmonic oscillator) is the Gaussian distribution.

If the median and quartiles are used as a given parameters instead of sample mean and variance, which are very sensitive to outlying observations, the family of distributions that are nonparametric analogue of Gaussian distribution and accounting for possible data asymmetry can be found [37].

**Conclusions**

Let us state a short summary.

The root density estimator is based on the representation of the probability density as a squared absolute value of a certain function, which is referred to as a psi function in analogy with quantum mechanics. The method proposed is an efficient tool to solve the basic problem of statistical data analysis, i.e., estimation of distribution density on the basis of experimental data.

The coefficients of the psi-function expansion in terms of orthonormal set of functions are estimated by the maximum likelihood method providing optimal asymptotic properties of the method (asymptotic unbiasedness, consistency, and asymptotic efficiency). An optimal number of harmonics in the expansion is appropriate to choose, on the basis of the compromise, between two opposite tendencies: the accuracy of the estimation of the function approximated by a finite series increases with increasing number of harmonics, however, the statistical noise level also increases.

The likelihood equation in the root density estimator method has a simple quasilinear structure and admits developing an effective fast-converging iteration procedure even in the case of multiparametric problems. It is shown that an optimal value of the iteration parameter should be found by the maximin strategy. The numerical implementation of the proposed algorithm is considered by the use of the set of Chebyshev-Hermite functions as a basis set of functions.

The introduction of the psi function allows one to represent the Fisher information matrix as well as statistical properties of the sate vector estimator in simple analytical forms. Basic objects of the theory (state vectors, information and covariance matrices etc.) become simple geometrical objects in the Hilbert space that are invariant with respect to unitary (orthogonal) transformations.

A new statistical characteristic, a confidence cone, is introduced instead of a standard confidence interval. The chi-square test is considered to test the hypotheses that the estimated vector equals to the state vector of general population and that both samples are homogeneous.

It is shown that it is convenient to analyze the sample populations (both homogeneous and inhomogeneous) using the density matrix.

The root density estimator may be applied to analyze the results of experiments with micro objects as a natural instrument to solve the inverse problem of quantum mechanics: estimation of psi function by the results of mutually complementing (according to Bohr) experiments. Generalization of the maximum likelihood principle to the case of statistical analysis of mutually complementing experiments is proposed. The principle of complementarity makes it possible to interpret the ill-posedness of the classical inverse problem of probability theory as a consequence of the lacking of the information from canonically conjugate probabilistic space.

The Fisher information matrix and covariance matrix are considered for a quantum statistical ensemble. It is shown that the constraints on the norm and energy are related to the gauge and time translation invariances. The constraint on the energy is shown to result in the suppression of high-frequency noise in a state vector approximated.

The analogy between the variational method in quantum mechanics and certain model problems of mathematical statistics is shown.

**References**

1. *A. N. Tikhonov and V. A. Arsenin.* Solutions of ill-posed problems. W.H. Winston. Washington D.C. 1977 .

2. *L. Devroye and L. Györfi*. Nonparametric Density Estimation: The $L_1$-View. John Wiley. New York. 1985.

3. *V.N. Vapnik and A.R. Stefanyuk*. Nonparametric methods for reconstructing probability densities Avtomatika i Telemekhanika 1978. Vol. 39. No. 8. P.38-52.

4. *V. N. Vapnik, T. G. Glazkova, V. A. Koscheev et al*. Algorithms for dependencies estimations. Nauka. Moscow. 1984 (in Russian).

5. *Yu. I. Bogdanov, N. A. Bogdanova, S. I. Zemtsovskii et al*. Statistical study of the time-to-failure of the gate dielectric under electrical stress conditions. Microelectronics. 1994. V. 23. N 1. P. 51 – 59. Translated from Mikroelektronika. 1994. V. 23. N1. P. 75-85.

6. *Yu. I. Bogdanov, N. A. Bogdanova, S. I. Zemtsovskii* Statistical modeling and analysis of data on time dependent breakdown in thin dielectric layers, Radiotekhnika i Electronika. 1995. N.12. P. 1874-1882.

7. *M. Rosenblatt* Remarks on some nonparametric estimates of a density function // Ann. Math. Statist. 1956. V.27. N3. P.832-837.

8. *E. Parzen* On the estimation of a probability density function and mode // Ann. Math. Statist. 1962. V.33. N3. P.1065-1076.

9. *E. A. Nadaraya* On Nonparametric Estimators of Probability Density and Regression, Teoriya Veroyatnostei i ee Primeneniya. 1965. V. 10. N. 1. P. 199-203.

10. *E. A. Nadaraya* Nonparametric Estimation of Probability Densities and Regression Curves. Kluwer Academic Publishers. Boston. 1989.

11. *J.S. Marron* An asymptotically efficient solution to the bandwidth problem of kernel density estimation. // Ann. Statist. 1985. V.13. №3. P.1011-1023.

12. *J.S. Marron* A Comparison of cross-validation techniques in density estimation // Ann. Statist. 1987. V.15. №1. P.152-162.

13. *B.U. Park, J.S. Marron* Comparison of data-driven bandwidth selectors // J. Amer. Statist. Assoc. 1990. V.85. №409. P.66-72.

14. *S.J. Sheather, M.C. Jones* A reliable data-based bandwidth selection method for kernel density estimation // J. Roy. Statist. Soc. B. 1991. V.53. №3. P.683-690.

15. *A. I. Orlov* Kernel Density Estimators in Arbitrary Spaces. in: Statistical Methods for Estimation and Testing Hypotheses. P. 68-75. Perm'. 1996 (in Russian).

16. *A. I. Orlov* Statistics of Nonnumerical Objects. Zavodskaya Laboratoriya. Diagnostika Materialov. 1990. V. 56. N. 3. P. 76-83.

17. *N. N. Chentsov (Čensov)* Evaluation of unknown distribution density based on observations. Doklady. 1962. V. 3. P.1559 - 1562.

18. *N. N. Chentsov (Čensov)* Statistical Decision Rules and Optimal Inference. Translations of Mathematical Monographs. American Mathematical Society. Providence. 1982 (Translated from Russian Edition. Nauka. Moscow. 1972).

19. *G.S. Watson* Density estimation by orthogonal series. Ann. Math. Statist. 1969. V.40. P.1496-1498.

20. *G. Walter* Properties of hermite series estimation of probability density. Ann. Statist. 1977. V.5. N6. P.1258-1264.

21. *G. Walter, J. Blum* Probability density estimation using delta sequences // Ann. Statist. 1979. V.7. №2. P. 328-340.

22. *H. Cramer* Mathematical Methods of Statistics, Princeton University Press, Princeton, 1946.

23. *A. V. Kryanev* Application of Modern Methods of Parametric and Nonparametric Statistics in Experimental Data Processing on Computers, MIPhI, Moscow, 1987 (in Russian).

24. *R.A. Fisher* On an absolute criterion for fitting frequency curves // Massager of Mathematics. 1912. V.41.P.155-160.

25. *R.A. Fisher* On mathematical foundation of theoretical statistics // Phil. Trans. Roy. Soc. (London). Ser. A. 1922. V.222. P. 309 – 369.

26. *M. Kendall and A. Stuart* The Advanced Theory of Statistics. Inference and Relationship. U.K. Charles Griffin. London. 1979.

27. *I. A. Ibragimov and R. Z. Has'minskii* Statistical Estimation: Asymptotic Theory. Springer. New York. 1981.

28. *S. A. Aivazyan and I. S. Enyukov, and L. D. Meshalkin* Applied Statistics: Bases of Modelling and Initial Data Processing. Finansy i Statistika. Moscow. 1983 (in Russian).

29. *L. D. Landau and E. M. Lifschitz* Quantum Mechanics (Non-Relativistic Theory). 3rd ed. Pergamon Press. Oxford. 1991.

30. *D. I. Blokhintsev* Principles of Quantum Mechanics, Allyn & Bacon, Boston, 1964.

31. *V. V. Balashov and V. K. Dolinov*. Quantum mechanics. Moscow University Press. Moscow. 1982 (in Russian).

32. *A. N. Tikhonov, A. B. Vasil`eva, A. G. Sveshnikov* Differential Equations. Springer-Verlag. Berlin. 1985.

33. *N. S. Bakhvalov, N.P. Zhidkov, G. M. Kobel'kov* Numerical Methods. Nauka. Moscow. 1987 (in Russian).

34. *N. N. Kalitkin* Numerical Methods. Nauka. Moscow. 1978 (in Russian).

35. *N. Bohr* Selected Scientific Papers in Two Volumes. Nauka. Moscow. 1971 (in Russian).

36. *P. J. Huber* Robust statistics. Wiley. New York. 1981.

37. *Yu. I. Bogdanov* Fisher Information and a Nonparametric Approximation of the Distribution Density// Industrial Laboratory. Diagnostics of Materials. 1998. V. 64. N 7. P. 472-477. Translated from Zavodskaya Laboratoriya. Diagnostika Materialov. 1998. V. 64. N. 7. P. 54-60.

38. *D. Bohm* A suggested interpretation of the quantum theory in terms of "hidden" variables. Part I and II // Phys. Rev. 1952. V.85. P.166-179 and 180-193

39. *I.J. Good, R.A. Gaskins* Nonparametric roughness penalties for probability densities // Biometrica. 1971. V.58. №2. P. 255-277.

40. *C. Gu, C. Qiu* Smoothing spline density estimation: Theory. // Ann. Statist. 1993. V. 21. №1. P. 217 – 234.

41. *P. Green* Penalized likelihood // in Encyclopedia of Statistical Sciences. Update V.2. John Wiley. 1998.

About the Author

Yurii Ivanovich **Bogdanov**

Graduated with honours from the Physics Department of Moscow State University in 1986. Finished his post-graduate work at the same department in 1989. Received his PhD Degree in physics and mathematics in 1990. Scientific interests include statistical methods in fundamental and engineering researches. Author of more than 40 scientific publications (free electron lasers, applied statistics, statistical modeling for semiconductor manufacture). At present he is the head of the Statistical Methods Laboratory (OAO "Angstrem", Moscow).

e-mail: bogdanov@angstrem.ru